\documentclass[useAMS,usenatbib]{mnras}
\usepackage{times}              	
\usepackage[fleqn]{amsmath}     
\usepackage{amssymb}
\usepackage{graphicx}
\usepackage{booktabs}
\usepackage[usenames, dvipsnames]{color}
\usepackage{rotating}
\usepackage{multirow}
\usepackage[normalem]{ulem}
\usepackage{myaasmacros}

\newcommand{\pfds}[1]{{{\bf \color{asparagus} PFdS: #1}}}

\newcommand{\ud}{\mathrm{d}}
\newcommand{\rhodm}{\rho_{\mathrm{DM},\odot}}
\newcommand{\rhob}{\rho_{\mathrm{b}}}
\newcommand{\vzsqmean}{\overline{v_z^2}}

\definecolor{asparagus}{rgb}{0.53, 0.66, 0.42}
\def\MWA{{\tt GSE60}}
\def\MWB{{\tt Sgr6}}
\def\MultiNest{{\tt MultiNest}}

\title[The local dark matter density in a wobbling disc]{Estimating the local dark matter density in a non-axisymmetric wobbling disc}

\author[Sivertsson et al.]{S. Sivertsson$^{1}$, J. I. Read$^{2}$, H. Silverwood$^{3}$, P. F. de Salas$^{1}$, K. Malhan$^{1}$, A. Widmark$^{1,4}$, \and C. F. P. Laporte$^{3,5}$, S. Garbari$^{6}$ and K. Freese$^{1,7,8}$.\\
$^1$ The Oskar Klein Centre for Cosmoparticle Physics, Department of Physics, Stockholm University, AlbaNova, SE-106 91 Stockholm, Sweden\\
$^2$ Department of Physics, University of Surrey, Guildford, GU2 7XH, Surrey, UK\\
$^3$ Institut de Ci\`encies del Cosmos (ICCUB), Universitat de Barcelona (IEEC-UB), Mart\`i Franqu\`es 1, E08028 Barcelona, Spain\\
$^4$ Dark Cosmology Centre, Niels Bohr Institute, University of Copenhagen, Jagtvej 128, 2200 Copenhagen N, Denmark \\
$^5$ Kavli Institute for the Physics and Mathematics of the Universe (WPI), The University of Tokyo Institutes for Advanced Study (UTIAS), The University of Tokyo, Chiba 277-8583, Japan\\
$^6$ Science Lab UZH, University of Zurich, Winterthurerstrasse 190, 8057 Zurich, Switzerland\\c
$^7$ Theory Group, Department of Physics, The University of Texas at Austin, 2515 Speedway, C1600, Austin, TX 78712-0264, USA\\
$^8$Nordic Institute for Theoretical Physics (NORDITA), Hannes Alfvéns v\"ag 12, 114 21 Stockholm, Sweden\\\\
}

\date{Accepted XXX. Received YYY; in original form ZZZ}
\pubyear{2021}

\begin{document}

\maketitle

\begin{abstract}
The density of dark matter near the Sun, $\rhodm$, is important for experiments hunting for dark matter particles in the laboratory, and for constraining the local shape of the Milky Way's dark matter halo. Estimates to date have typically assumed that the Milky Way's stellar disc is axisymmetric and in a  steady-state. Yet the Milky Way disc is neither, exhibiting prominent spiral arms and a bar, and vertical and radial oscillations. We assess the impact of these assumptions
on determinations of $\rhodm$ by applying a free-form, steady-state, Jeans method to two different $N$-body simulations of Milky Way-like galaxies. In one, the galaxy has experienced an ancient major merger, similar to the hypothesized Gaia-Sausage-Enceladus; in the other, the galaxy is perturbed more recently by the repeated passage and slow merger of a Sagittarius-like dwarf galaxy. We assess the impact of each of the terms in the Jeans-Poisson equations on our ability to correctly extract $\rhodm$ from the simulated data. We find that common approximations employed in the literature -- axisymmetry and a locally flat rotation curve -- can lead to significant systematic errors of up to a factor $\sim 1.5$ in the recovered surface mass density $\sim 2$\,kpc above the disc plane, implying a fractional error on $\rhodm$ of order unity. However,  once we add in the tilt term and the rotation curve term in our models, we obtain an unbiased estimate of $\rhodm$, consistent with the true value within our 95\% confidence intervals  for realistic 20\% uncertainties on the baryonic surface density of the disc. Other terms --- the axial tilt, 2:nd Poisson and time dependent terms --- contribute less than 10\% to $\rhodm$ (given current data) and can be safely neglected for now. In the future, as more data become available, these terms will need to be included in the analysis.
\end{abstract}

\section{Introduction}\label{sec:intro}

The local dark matter density, $\rho_{\rm DM, \odot}$, is an estimate of the density of dark matter (DM) near the Sun, typically averaged over a small volume of width $\sim 0.2-1$\,kpc and height $\sim 0.2-3$\,kpc \citep[e.g.][]{Read:2014qva}. In combination with other Galactic information, including the rotation curve, it yields an estimate of local shape of the Milky Way's dark matter halo that can be used to test alternative gravity models and galaxy formation theories \citep[e.g.][]{Read:2014qva}. It is also an important input in the quest to discover particle dark matter \citep[e.g.][]{Bertone:2004pz,Roszkowski:2017nbc}. The expected signal strength in many experimental searches, such as direct detection experiments searching for elastic scattering events, directly depends on the value of $\rho_{\rm DM, \odot}$. An accurate estimate is, therefore, needed to measure particle properties in the event of a detection.

\begin{figure}
\includegraphics[width=\columnwidth]{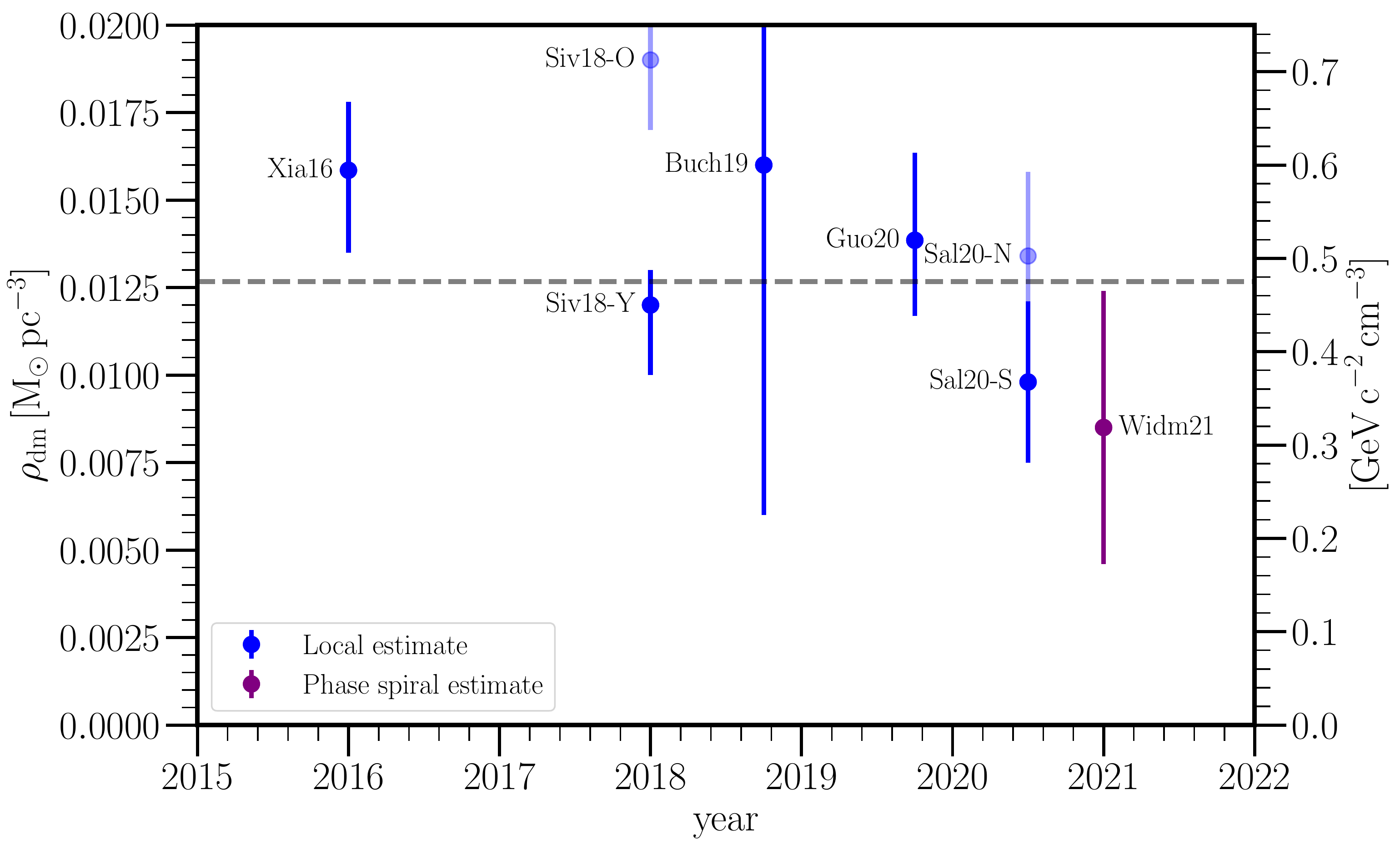}   
\caption{Recent local estimates of the local dark matter density, $\rhodm$. The mean of all data points is shown by the horizontal dashed line. At first sight, the agreement seems to be excellent as most data points agree within in their 68\% confidence intervals (error bars, as marked). However, there remain potentially significant sources of systematic error. The light blue points show alternative, disfavoured, measurements of $\rhodm$ using different tracer stars. In the case of \citet{Sivertsson:2017rkp}, these are `$\alpha$-old' stars that are kinematically hotter; for \citet{Salomon20}, these are stars in the Northern Hemisphere that show more prominent signs of disequilibrium than those in the Southern Hemisphere. Finally, the purple datapoint shows a recent determination from \citet{Widmark21} that is distinct from the other local methods in that it harnesses the fact that the Milky Way disc is wobbling and ringing. (Data taken from \citet{Xia:2015agz,Sivertsson:2017rkp,Buch:2018qdr,Guo2020,Salomon20,Widmark21}.).}
\label{fig:rhodmsummary}
\end{figure}

Recent estimates of $\rho_{\rm DM, \odot}$ make use of a variety of methods \citep[e.g.][]{deSalas:2020hbh}. `Local' methods use Jeans and/or distribution function models over small volumes near the Sun \citep[e.g.][]{Xia:2015agz,Sivertsson:2017rkp,Guo2020,Salomon20}. `Global' methods use stellar tracers over larger volumes \citep[e.g.][]{Cole:2016gzv,Wegg19}, or use the Milky Way's circular velocity, assuming that its dark matter halo is spherically symmetric \citep[e.g.][]{Benito:2019ngh,Karukes:2019jxv,deSalas:2019pee}. More recently, newer methods have been introduced, like `made to measure' \citep[e.g.][]{Syer96,Bovy18}, and a recent attempt to harness the Milky Way disc's disequilibrium to estimate $\rho_{\rm DM, \odot}$ \citep{Widmark21}. Some studies exploit the vertical kinematics of stars up to a height $z \sim 1-3\,\mathrm{kpc}$ above the Galactic disc \citep[e.g.][]{Xia:2015agz,Sivertsson:2017rkp,Guo2020,Salomon20}, while others use a similar methodology much closer to the plane ($<200$\,pc; \citealt{Schutz:2017tfp,Buch:2018qdr,Widmark:2018ylf}). The former has the advantage that dark matter makes up a larger fraction of the mass at these heights, but the disadvantage that it is less local. Recently, \cite{Cole:2016gzv} have fit a global distribution function model to a wide range of data for the Milky Way, while \citet{Wegg19} fit the azimuthally averaged Jeans equations to a sample of $\sim 16,000$ RR Lyrae stars from Gaia DR2 over a Galactocentric distance range of $1.5 < R < 20$\,kpc. A review of historic estimates of $\rhodm$ is given in \citet{Read:2014qva}, and more recent estimates are reviewed in \citet{deSalas:2020hbh}; we summarise recent local estimates in Figure \ref{fig:rhodmsummary}\footnote{Note that these studies all assume a very similar model for the baryonic surface density, based on \citet{McKee:2015hwa} and \citet{flynn06}. As such, we need not worry about additional scatter between the data points due to the known degeneracy between the baryonic surface density of the disc, $\Sigma_b$, and $\rhodm$ \citep{Sivertsson:2017rkp}. The one exception to this is \citet{Salomon20} who allow $\Sigma_b$ to vary freely in their fits. However, they derive $\rhodm$ from a subtantially higher height of $\sim 3$\,kpc at which they are no longer very sensitive to $\Sigma_b$.}.

Most of the above estimates of $\rhodm$ assume a steady state pseudo-equilibrium such that the distribution function of the tracers stars has no explicit time dependence; most additionally assume axisymmetry. We have known for a long time that the Milky Way is not axisymmetric, displaying prominent spiral arms and a bar \citep[e.g.][]{Hou14,Camargo15,Hilmi20}. Furthermore, with the advent of large surveys like the RAdial Velocity Experiment and the Sloan Digital Sky Survey, we have discovered that the Milky Way disc is `wobbling', with significant vertical and radial oscillations in density and velocity \citep{Widrow12,Williams13}. The ESA-Gaia satellite has brought this into sharp focus \citep{schoenrich18,Bennett2019}. A major discovery that has come from its second data release (DR2) is the detection of `phase-space spiral' in the local stellar disc \citep{Antoja2018}. Several authors have provided potential explanations for this feature \citep[see e.g.][]{Laporte2018, Antoja2018, Binney18, Laporte2019, Bland-Hawthorn2019,Khoperskov:1811.09205}. When dissecting the phase-space spiral in stellar ages, \citet{Laporte2019} detect it at all stellar ages down to very young ages $\sim0.5-1\,\rm{Gyr}$ pointing towards a recent perturbation (as opposed to bar buckling). Taken at face value, this is consistent with orbital timescales of Sagittarius \citep{johnston05,penarrubia10} and those derived for phase-mixing timescales through toy-modelling in the original study of \citet{Antoja2018}, favouring the impact scenario \citep{Laporte2018}. This is currently the only mechanism able to simultaneously reconcile perturbations on large and small scales in the disc \citep{Slater14, price-whelan15,schoenrich18,deboer18, bergemann18,laporte20}. Although the details of the perturbation remain to be worked out, most authors agree that some form of disequilibrium (i.e. a non-steady-state disc) is at work in the solar neighbourhood and across the Galaxy both vertically and radially \citep{minchev09,gomez13,monari16,Laporte2018,Hunt2018, Antoja2018, laporte20b,Lopez-Corredoira2020,xu20}. 


Given that the Milky Way disc is not axisymmetric nor in a steady state, we can reasonably ask how secure recent estimates of $\rhodm$ are. Early work demonstrated that Jeans methods, that assume an axisymmetric potential and separable radial and vertical stellar motions, are able to obtain a statistically unbiased estimate of $\rhodm$ when applied to a few thousand tracer stars, even in the face of prominent spiral arms and a bar \citep{Garbari2011}. However, the Gaia data presents us with {\it millions} of tracer stars that extend much higher above the disc \citep[e.g.][]{Read:2014qva}. For this quality and extent of data, systematic errors will almost certainly dominate the error budget, especially if the Milky Way's disc lies much further from a pseudo-equilibrium state than was assumed in \citet{Garbari2011}. More recently, \citet{Haines2019} have applied an axisymmetric Jeans method to mock data from an $N$-body simulation of the Milky Way that includes a recent encounter with a Sagittarius-like dwarf galaxy \citep{Laporte2018}. They find that, in this case, estimates of $\rhodm$ can become systematically biased by up to a factor of $\sim 1.5$, which is substantially larger than the formal uncertainties on $\rhodm$. Indeed, this could explain systematic variances in $\rhodm$ reported in the literature for estimates derived from different stellar tracer populations in the disc, and between red clump stars in the Northern and Southern Hemispheres (see \citealt{Sivertsson:2017rkp,Salomon20} and Figure \ref{fig:rhodmsummary}).

To address the above issues, in this paper we assess how well a steady-state free-form (non-parametric)\footnotemark\ Jeans method can recover $\rhodm$ from a non-axisymmetric, non-steady state disc. We further develop the mass modelling method introduced in \citet{Silverwood:2015hxa} and \citet{Sivertsson:2017rkp} to generalise the functional form of the tracer density and velocity anisotropy profiles, and to remove any assumption of axisymmetry. We then apply this method to mock data generated from a simulated `wobbling' disc, seeded by either an ancient massive merger or a more recent Sagittarius-like merger (this latter simulation is identical to the one used in \citealt{Haines2019}). We assume a `best-case' scenario of zero observational uncertainties, a volume complete sample of tracer stars, and perfect measurement of all steady-state terms in the Jeans equation. This allows us to isolate the impact of common modelling assumptions on our recovery of $\rhodm$, carefully separating out bias induced by disequilibrium from bias due to other common assumptions, like axisymmetry. More realistic mocks should include the impact of observational uncertainties, survey selection effects and the current impact of the Large Magellanic Cloud that is believed to be on its first infall and about a fifth of the mass of the Milky Way \citep[e.g.][]{weinberg98,besla07,gomez15,penarrubia16,Laporte18a,Laporte18b,erkal19,read19,GaravitoCamargo19,donaldson21}. We will consider these additional complications in future work.

\footnotetext{``Free form'' or ``non-parametric'' (often used interchangeably), means a model in which there are more (typically many more) parameters than data constraints. Such a model is deliberately under-constrained and must be fit using a method like Markov Chain Monte Carlo that allows parameter degeneracies to be explored and marginalised over. In this way, we obtain confidence intervals of some quantity of interest -- in this case the local dark matter density -- that are hopefully not sensitive to potentially arbitrary model choices (for further discussion on this, see e.g. \citealt{Coles14} and \citealt{ReadSteger17}).}

This paper is organised as follows. In \S\ref{sec:jeans}, we describe our mass modelling method that builds on the free-form method presented in \citet{Silverwood:2015hxa} and \citet{Sivertsson:2017rkp}. In \S\ref{sec:simulations}, we describe the $N$-body simulations that we use in this work, and how we generate our mock data from these simulations. In \S\ref{sec:results}, we present the results from applying our method to the mocks. In \S\ref{sec:discussion}, we discuss our results and their broader implications for estimating $\rhodm$. Finally, in \S\ref{sec:conclusions}, we present our conclusions.

\section{Simulations} \label{sec:simulations}

\subsection{Description of the simulations}

We test our free-form Jeans method (\S\ref{sec:jeans}) on two $N$-body simulations of MW-like galaxies, both of which are perturbed by an external, smaller, galaxy. One of the simulations -- \MWA\ -- is an $N$-body simulation of a MW-like galaxy that experienced a $M\sim 10^{11}$\,M$_\odot$ merger $\sim 4$ Gyrs ago, and then evolved without any further external perturbations \citep{Garbari_thesis}. In this simulation, the MW-like galaxy is identical to that used in \citet{Garbari2011}, which is set up as described in \citet{widrow05} (see Table \ref{tab:sims} for a summary of its properties). The motivation for this case stems from the recent analysis of the local stellar halo which shows that more than half of the $\rm{[Fe/H]<-1\,dex}$ stars appear to originate from a single massive accreted system -- the `Gaia-Sausage-Enceladus' (GSE) satellite galaxy (e.g., \citealt{BelokurovGSE2018, HelmiGSE2018, Haywood2018, Mackereth2019}) that had a mass of $\sim 1.0\times 10^{11}M_\odot$ ($\sim 1/10$ total mass ratio). We will refer to this simulation as \MWA, where ``60'' refers to the initial orbit inclination, in degrees, of the merger with respect to the main galaxy's disc plane \citep{Garbari_thesis}.

\begin{table}
    \centering
    \resizebox{0.475\textwidth}{!}{
    \begin{tabular}{l | c c c c c | l}
    \hline
     & $N$ & $M$ & $\epsilon$ & $R_{1/2}$ & $z_{1/2}$ & \\
     & $[10^6]$ & $[10^{10}\,{\rm M}_\odot]$ & $[{\rm kpc}]$ & $[{\rm kpc}]$ & $[{\rm kpc}]$ & Notes \\
     \hline
     \MWA\ & & & & & & 
     
     \multirow{3}{*}{\parbox{1.5cm}{
     ${\sim}10^{11}$\,M$_\odot$ satellite merger 4\,Gyrs ago}} 
     
     \\
    \cline{1-1}
 
    Disc & 30 & 5.30 & 0.015 & 4.99 & 0.17 & \\
    Bulge & 0.5 & 0.83 & 0.012 & -- & -- & \\
    Halo & 15 & 45.40 & 0.045 & -- & -- & \\
    \hline
    \MWB\ & & & & & & 
     
     \multirow{3}{*}{\parbox{1.5cm}{${0.8\times} 10^{11}$\,M$_\odot$ merger, ongoing}} 
     
     \\
     \cline{1-1}
      
    Disc & 5 & 6.0 & 0.03 & 5.88 & 0.37 & \\
    Bulge & 1 & 1.0 & 0.03 & -- & -- & \\
    Halo & 40 & 100 & 0.06 & -- & -- & \\
    \hline
    \end{tabular}
    }
    \caption{The $N$-body simulations used to generate mock data. From left to right, the columns give: the simulation label; number of particles, mass and force softening of each component; the half mass scale length and height of the disc prior to the satellite merger; and some notes on each simulation (see \S\ref{sec:simulations} for details).     }
    \label{tab:sims}
\end{table}

Our main emphasis, however, will be on the second $N$-body simulation -- \MWB\ -- described in \cite{Laporte2018}, set up using {\sc galic} \citep{Yurin14}. In this simulation, a Sagittarius-like dwarf galaxy ($M_{200} \sim 0.6\times 10^{11}M_\odot$) is accreted onto a previously unperturbed Milky Way-like galaxy (see Table \ref{tab:sims} for a summary of its properties). Following the Gaia DR2 release, \citet{Laporte2019} re-analysed this $N$-body simulation, finding that the repeated impacts of the orbiting Sagittarius-like dwarf galaxy excites oscillations in the disc that reproduced many of the newly uncovered features in the Gaia DR2 data \citep{Katz2018}. We will refer to this simulation as \MWB, where ``6'' refers to the snapshot location in time, $t=6$\,Gyr from the start of the simulation, which is the same snapshot as used in \citet{Haines2019}. This time was not chosen with intention to use a ``present-day'' snapshot for the MW-Sgr-like system, but rather specifically pick a time shortly following a pericentric passage by a satellite. Note that we also applied our Jeans analyses method to earlier snapshots in the simulation, for which the MW-like galaxy had not yet been perturbed. Our method worked very well on this unperturbed disc and so, for brevity, we do not discuss those results here.

\begin{figure}
\includegraphics[width=\columnwidth, trim={0.4cm 0.5cm 0.4cm 0.4cm},clip]{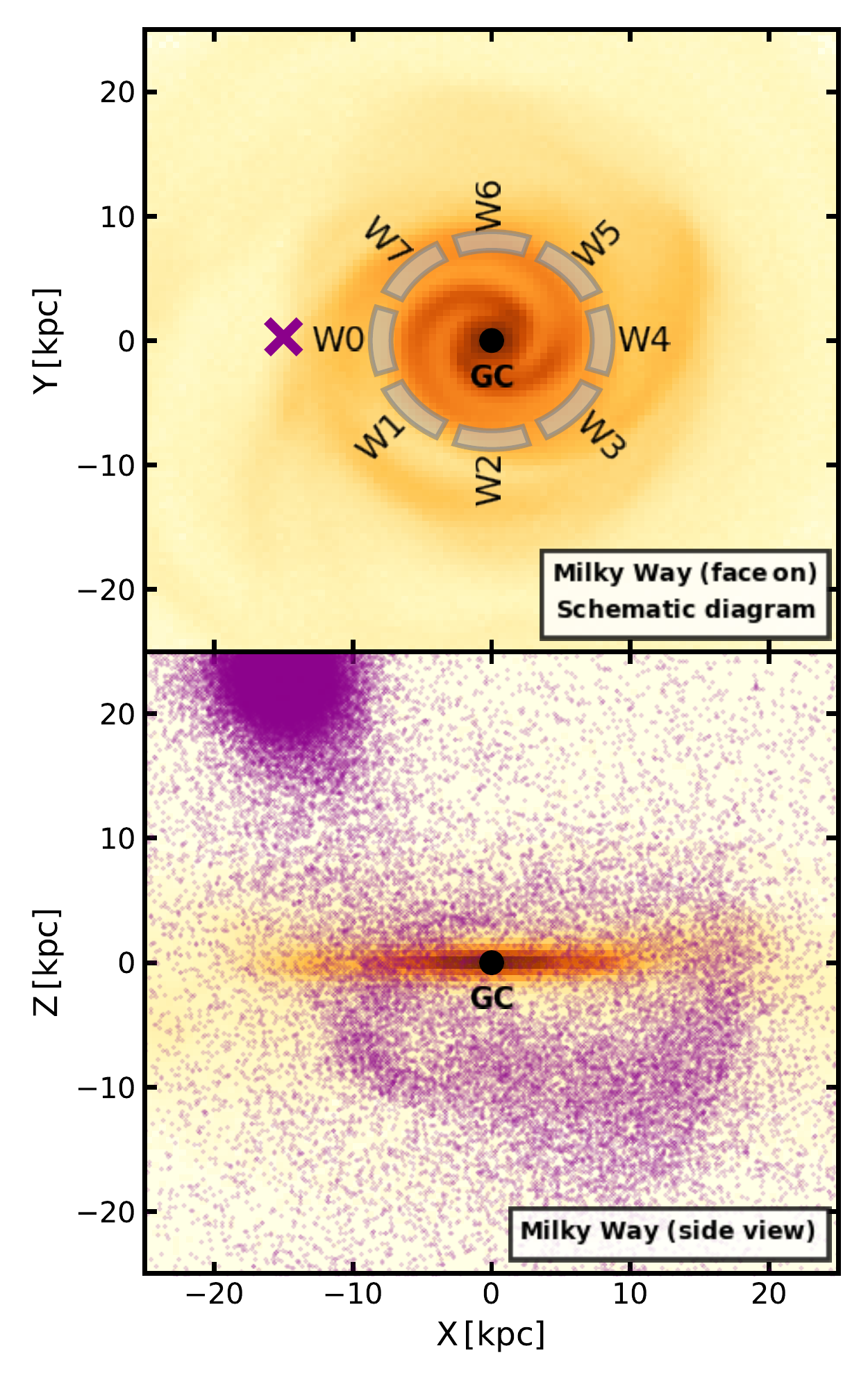}   
\caption{Face-on (top) and edge-on (bottom) view of the \MWB\ simulation at $t=6$\,Gyrs from \citet{Laporte2018}, showing the ongoing merger of a Sagittarius-like dwarf (purple cross and dots) with an MW-like host galaxy (yellow-orange). The top panel shows the location and extent of the 8 wedges in the disc plane from which we extract our mock data.}
\label{stellar_locations}
\end{figure}

The two simulations investigated are complementary. The \MWA\ simulation shows a MW-like galaxy that has undergone an ancient but violent merger from which the system has since had time to relax. The \MWB\ simulation, on the other hand, considers the impact of an ongoing perturbation of the MW-like galaxy. A visual impression of the \MWB\ simulation at $t=6$\,Gyrs is shown in Fig.~\ref{stellar_locations}, where the top and bottom panels show a face-on and edge-on view of the stellar disc, respectively. In this snapshot, the remnant of the Sagittarius-like dwarf (purple cross and dots) is moving downwards, towards the MW-like disc. For a visual impression of the \MWA\ simulation (that has no ongoing merger), see \citet{Garbari2011}.

\subsection{Generating the mock data} \label{sec:mocks}

When analysing the N-body simulations, we divide the galactic disc into 8 `wedges', with a small gap between each wedge, as shown in Fig.~\ref{stellar_locations}. We analyse each wedge separately, essentially imagining 8 different locations in the simulated galaxy where the Sun might be placed. Each wedge spans a radial range $7.5<R<8.5$\,kpc from the galactic centre, subtends an angle of $\pi/5$, and is centered around $\frac{\pi}{4}N-\pi$, where $N\in[0,7]$ is the number of the wedge, as shown in Fig.~\ref{stellar_locations}. The wedges used are identical for both the \MWA\ and \MWB\ simulations.

When generating the mock data from each wedge, the data are binned using equidistant bins in height, $z$. We assume `ideal data' with zero measurement uncertainty, volume complete tracer stars, and consistent photometric and kinematic tracer populations (see e.g. the discussion on this in \citealt{Read:2014qva}). This will allow us to focus on how common modelling assumptions drive systematic errors on $\rhodm$. Furthermore, the tracer stars only include stars which originated in the Milky Way-like galaxy. This assumes that these can be observationally separated from the accreted stars using, for example, their distinct chemistry \citep[e.g.][]{Ruchti2015}. We leave an exploration of more realistic uncertainties, selection functions and inconsistency to future work.

The wedges in \MWA\ and \MWB\ have 20,000-48,000 and 100,000-170,000 stars within $\pm $2\,kpc of z=0, respectively. For Gaia, we can expect millions of tracers near the Sun \citep{Read:2014qva}, but after quality, magnitude and colour cuts, and avoiding the high-extinction disc, this number is dramatically reduced. For example, \citet{Salomon20} present a recent estimate of $\rhodm$ from Gaia DR2 red clump stars; after such cuts, their total sample is $\sim 40,000$ stars, with half in the Northern and half in the Southern Hemisphere. As such \MWA\ and \MWB\ are indicative of the current Gaia sampling-error.

\section{The mass modelling method}\label{sec:jeans}

\subsection{The Jeans-Poisson equations}
Our mass modelling method builds on the free-form Jeans method presented in \citet{Silverwood:2015hxa} and \citet{Sivertsson:2017rkp}. To derive $\rhodm$, we solve the vertical Jeans equations \citep{Jeans1915,Binney_Tremaine_Book} locally for each wedge in cylindrical coordinates, $(R,\phi,z)$:

\begin{eqnarray}
\underbrace{\frac{1}{R\nu}\frac{\partial}{\partial R}\left( R\nu\mathrm{ \ }\overline{v_Rv_z} \right)}_{\mathrm{tilt \ term: \ }\mathcal{T}} +
\underbrace{\frac{1}{R\nu}\frac{\partial}{\partial \phi}\left(\nu\mathrm{ \ }\overline{v_\phi v_z} \right)}_{\mathrm{axial \ tilt \ term: \mathcal{A}}} +
\frac{1}{\nu}\frac{\partial}{\partial z}\left(\nu\mathrm{ \ }\overline{v_z^2} \right) + \nonumber  \\ 
+ \underbrace{\frac{1}{\nu}\frac{\partial}{\partial t}\left(\nu \mathrm{ \ }\overline{v_z} \right)}_{\mathrm{explicit \ time \ dependent \ term}} + \frac{\partial \Phi}{\partial z} = 0,
\label{Jeans_eq}
\end{eqnarray}
where $\nu$ is the spatial number density of tracer stars, $\Phi$ is the gravitational potential, and $\overline{v_{i}}$ and $\overline{v_{ij}}$ are the mean velocity and velocity dispersion tensor (where $i,j=R,\phi,z$), respectively.

The Poisson equation links the gravitational potential to the total matter density, $\rho(R,\phi,z)$, which in cylindrical coordinates is given by:
\begin{equation}
\underbrace{\frac{1}{R}\frac{\partial}{\partial R}\left(R\frac{\partial\Phi}{\partial R}\right)}_{\mathrm{rot \ curve \ term: \ }\mathcal{R}}+
\underbrace{\frac{1}{R^2}\frac{\partial^2 \Phi}{\partial \phi^2}}_{\mathrm{2:nd \ Poisson \ term}}+
\frac{\partial^2 \Phi}{\partial z^2} = 4\pi G\rho ,
\label{Poisson_eq}
\end{equation}
where $G$ is Newton's constant. Notice that, at this stage, Eqs.~(\ref{Jeans_eq}) and (\ref{Poisson_eq}) have not introduced any assumptions about symmetry or steady-state pseudo-equilibrium. If the system is axisymmetric, then the axial tilt term and 2nd Poisson term will vanish; if steady-state, then the explicit time dependent term will vanish.

Given an observed density of tracer stars, $\nu(R,\phi,z)$ and their velocity moments $\overline{v_{i}}$ and $\overline{v_{ij}}$, we can, at least in principle, solve Eqs. (\ref{Jeans_eq}) and (\ref{Poisson_eq}) to determine the matter density, $\rho(R,\phi,z)$. Subtracting off the contribution from visible stars, gas and stellar remnants yields an estimate of the mean dark matter density in each wedge, $\rhodm$ \citep[e.g.][]{Read:2014qva}.

We now discuss each of the terms in Eqs. (\ref{Jeans_eq}) and (\ref{Poisson_eq}), common approximations in the literature that neglect some subset of these terms, and their likely importance for the determination of $\rhodm$.

\subsection{The 1D approximation}\label{sec:1dapprox}
If we assume for the Galaxy: 1) a mass distribution close to axysimmetric; 2) a steady-state; 3) approximately separable radial and vertical stellar motions; and 4) an approximately flat rotation curve, then the Jeans-Poisson equations become:
\begin{equation}
\frac{1}{\nu}\frac{\partial}{\partial z}\left(\nu\mathrm{ \ }\overline{v_z^2} \right) + \frac{\partial \Phi}{\partial z} = 0, \ \ \frac{\partial^2 \Phi}{\partial z^2} = 4\pi G\rho(z).
\label{simple_Jeans_Poisson}
\end{equation}
which is known as the `1D approximation', since only motion in $z$ remains in the equations \citep{Read:2014qva}.

Eqs. (\ref{simple_Jeans_Poisson}) can be solved in this differential form \citep[e.g.][]{Garbari2012}, but this requires an estimate of the derivative of the tracer density, $\nu(z)$, and the vertical velocity dispersion, $\overline{v_z^2}(z)$ that can be quite noisy. For this reason, we favour integrating over $z$ to give \citep{Silverwood:2015hxa}:
\begin{equation}
    \overline{v_z^2}(r) = \frac{1}{\nu(z)}\int_z^\infty \nu(z') K_z(z') dz' + \frac{C}{\nu(z)}
    \label{eqn:1dapprox}
\end{equation}
where:
\begin{equation}
K_z = \frac{\partial \Phi}{\partial z} = 4\pi G\left(\int_0^z \rho(z') \ud z'\right) = 2\pi G \Sigma(z)
\label{eqn:Kz}
\end{equation} 
is the `vertical force'; 
$C\propto \partial\Phi/\partial z|_{z=0}$ is a normalisation constant that we neglect (see \citealt{Sivertsson:2017rkp});
and $\Sigma(z)$ is the disc surface density.  Note that this equation is only strictly correct if the matter distribution is symmetric around $z=0$, for which $K_z(0) = 0$. For the mock data we consider here, even when there are north-south asymmetries, $K_z(0)$ is sufficiently small as to be negligible. As such, in this paper we can meaningfully apply this method separately to $z > 0$ and $z < 0$ data samples. We note that we define $z=0$ by requiring the total mass of tracer stars within approximately two scale heights in the $z>0$ direction to equal that of the $z<0$ direction.

Most work to date on estimating $\rhodm$ has employed something akin to the 1D approximation \citep[e.g.][]{Read:2014qva,Xia:2015agz,Sivertsson:2017rkp,Guo2020,Salomon20}. Here, we will also use this approximation as our default method. However, we will then add in each of the missing terms from the Jeans equation one at a time to assess their importance. In all cases, we will assume a perfect measurement of each term as calculated directly from the simulations. This will allow us to determine which terms are important and which can be discarded and, in particular, whether we must include time dependent terms in order to obtain a robust estimate of $\rhodm$. We now consider each of the terms neglected in the above 1D approximation in turn. 

\subsection{The rotation curve term}\label{sec:rotcurve}

We can estimate the impact of the rotation curve term, $\mathcal{R}$, on $\rhodm$, as a systematic error of size:

\begin{equation}
    \Delta\rho = \frac{\mathcal{R}}{4\pi G}      = \frac{1}{4\pi G R}\frac{\partial v_c^2}{\partial R}
    \label{eqn:rotcurveterm}
\end{equation}
where $v_c^2(R) = R \partial \Phi / \partial R$ gives the rotation speed of circular orbits in the $z=0$ plane. Thus, if the rotation curve is exactly flat, then $\mathcal{R}$ vanishes and $\Delta\rho=0$.

\begin{figure}
\includegraphics[width=\columnwidth, trim={0.2cm 0.2cm 0.2cm 0.2cm},clip]{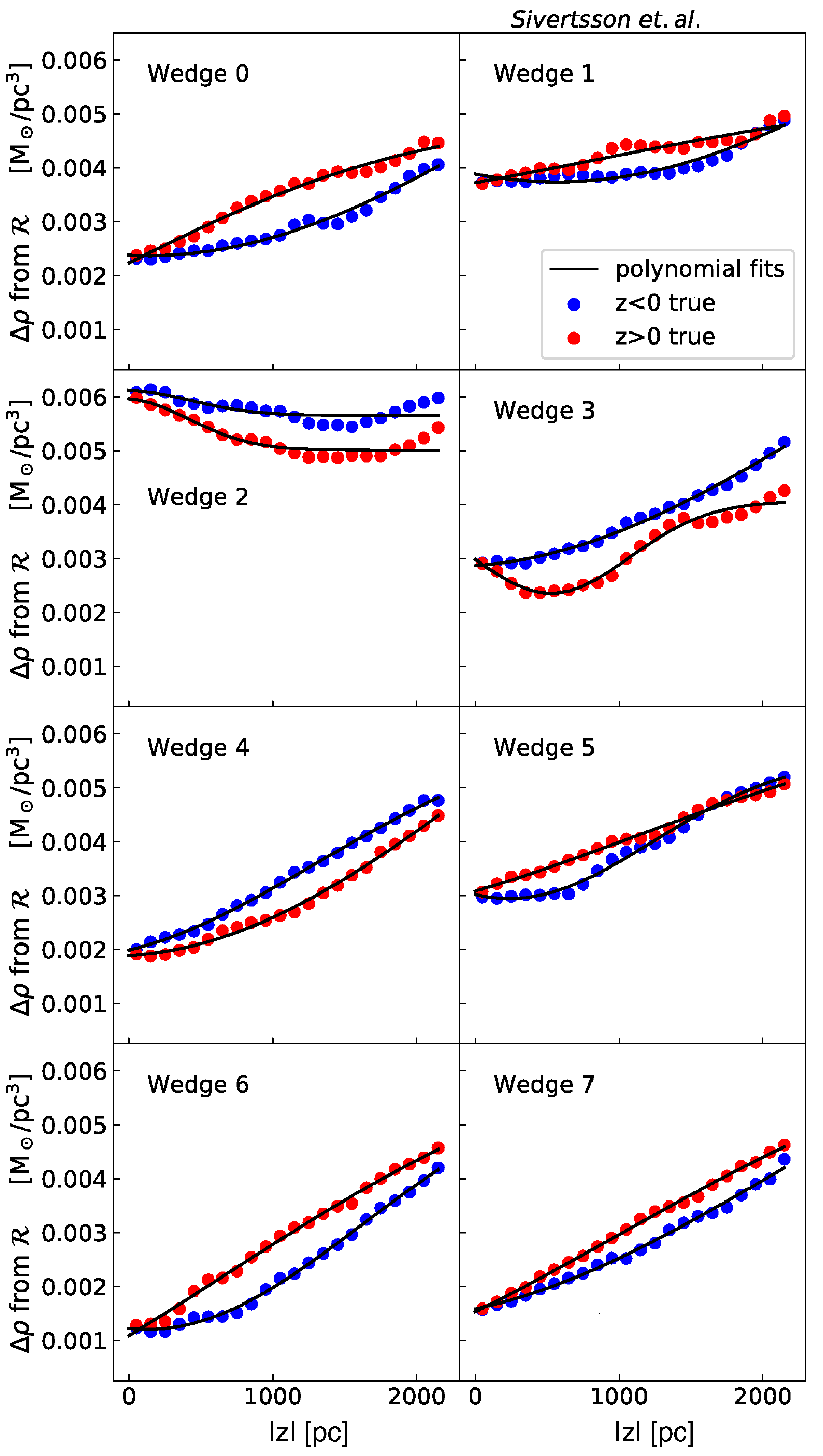}
\vspace{-4mm}
\caption{The rotation curve term contribution to the recovered local density for the \MWA\ simulation as a function of $|z|$ for the $z>0$ (red) and $z<0$ (blue) directions. The black lines are fits of the form shown in Eq.~\eqref{rot_term_fit}. Notice that in many of the wedges, $|\Delta \rho|$ approaches about half of the local dark matter density, $\rhodm \sim 0.01M_\odot/$pc$^2$. As such, this is an important term that should not be neglected.}
\label{rot_curve_term_60}
\vspace{-4mm}
\end{figure}

\begin{figure}
\includegraphics[width=\columnwidth]{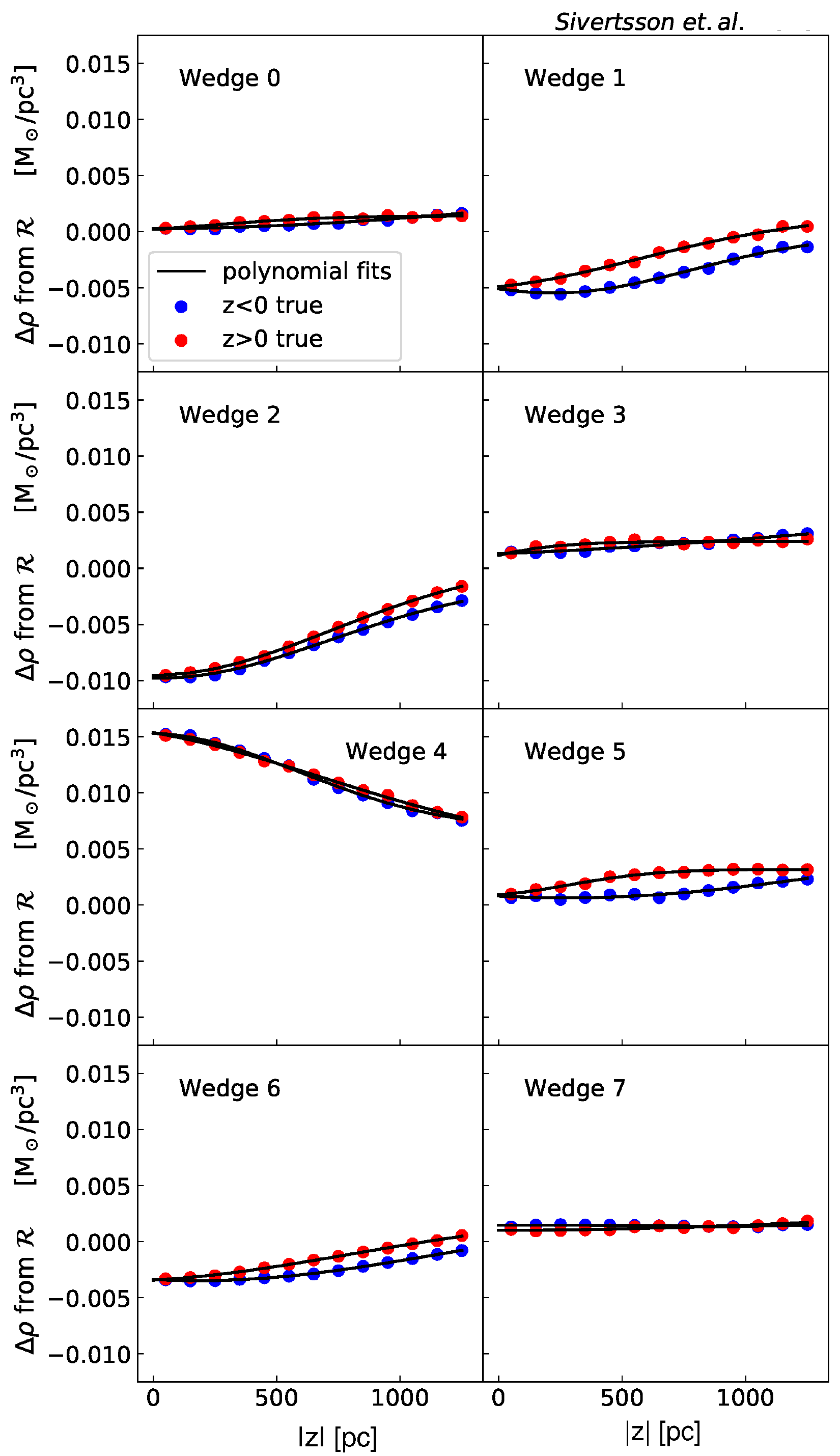}
\vspace{-4mm}
\caption{As Fig.~\ref{rot_curve_term_60} but for simulation \MWB. Notice that for this simulation, in which the satellite interaction is still ongoing, in several wedges $|\Delta \rho|$ is now of the order of the local dark matter density, $\rhodm \sim 0.01M_\odot/$pc$^2$. As with \MWA, this is an important term that should not be neglected.}
\vspace{-4mm}
\label{rot_curve_term_Sgr600}
\end{figure}

In Figs.~\ref{rot_curve_term_60} and \ref{rot_curve_term_Sgr600}, we plot $\Delta \rho$ for each wedge in the \MWA\ and \MWB\ simulations, respectively. We calculate $\Delta \rho$ directly from the known gravitational potential of the simulations, averaging over the size of the wedge weighted by the local tracer density. For the real Milky Way, we can estimate $|\Delta \rho|$ from the radial derivative of the local rotation curve as a function of height above the plane \citep[e.g.][]{Garbari2012,Li19}. We discuss this further in \S\ref{sec:discussion}.

As can be seen, across all wedges and both simulations $|\Delta \rho|$ can often be as large as $\sim$half of the local dark matter density. It is positive for all wedges in the \MWA\ simulation, implying that one would underestimate the true underlying surface density if the $\mathcal{R}$ term is neglected. By contrast, for the \MWB\ simulation, $\Delta \rho$ varies strongly between the wedges and can be positive or negative -- an indication of significant non-axisymmetry. For wedge 4 in \MWB, $\Delta \rho$ is comparable to $\rhodm$ and so neglecting it in the modelling will be a particularly poor approximation.

In \S\ref{sec:results_rotcurve}, we include $\Delta\rho(z)$ in our mass model as a correction to Eq.~\eqref{eqn:Kz}, modelling it as an analytic function of the form:

\begin{equation} \label{rot_term_fit}
\Delta\rho \simeq A + B\mathrm{e}^{-\left(\frac{z-m}{s}\right)^2},
\end{equation}
where $A$, $B$, $m$ and $s$ are fitted to the $\mathcal R/4\pi G$ data. These fits are shown as black lines in Figs.~\ref{rot_curve_term_60} and \ref{rot_curve_term_Sgr600}.

\subsection{The 2:nd Poisson term}\label{sec:2nd-poisson-term}
\begin{figure}
\includegraphics[width=\columnwidth, trim={0.5cm 0.5cm 0.4cm 0.0cm},clip]{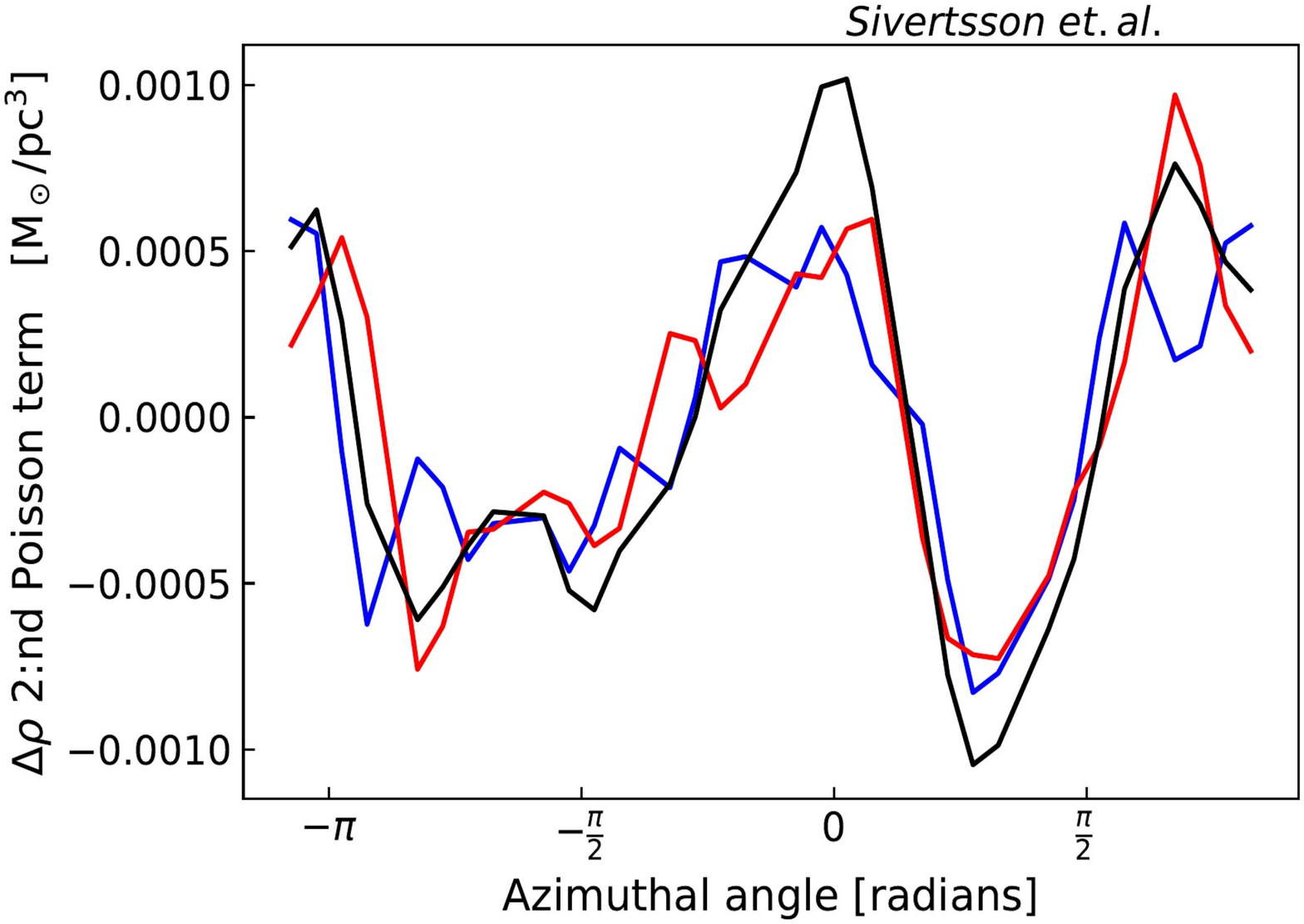}
\caption{The 2:nd Poisson term as a function of azimuthal angle, $\phi$, averaged over the radial interval $7.5<R<8.5$\,kpc, for the \MWB\ simulation. The black line corresponds to $z=0$, while the blue and red lines correspond to $z=-1$\,kpc and $z=1$\,kpc, respectively. This term is everywhere less than 10\% of $\rhodm$ (also for the \MWA\ simulation) and so can be safely neglected.}
\label{2nd_Poisson_term_Sgr600}
\end{figure}

Similarly to the rotation curve term, we can estimate the impact of the 2:nd Poisson term (Eq.~\ref{Poisson_eq}) on $\rhodm$, as a systematic error of size:

\begin{equation}
\Delta \rho_{\rm 2:nd} = \frac{1}{4 \pi G R^2}\frac{\partial^2 \Phi}{\partial \phi^2}.
\label{2nd_Poisson_term}
\end{equation}
 In Fig.~\ref{2nd_Poisson_term_Sgr600} we show $\Delta \rho_{\rm 2:nd}$ for the \MWB\ simulation, calculated similarly to the rotation curve term, averaging over the radial interval $7.5<R<8.5$\,kpc, but no longer averaging over azimuth. As can be seen, $|\Delta \rho_2| < 0.001~$M$_\odot/$pc$^3$, which corresponds to about 10\% of $\rhodm$. Averaging over the wedges (that each span a range of azimuth) will further dilute the importance of $\Delta \rho_2$ and so we can safely neglect this term. (The 2:nd Poisson term is also very small for the \MWA\ simulation and so we omit this plot for brevity.)
 
\subsection{The tilt term}\label{sec:tilt}
Following \citet{Silverwood:2015hxa} and \citet{Sivertsson:2017rkp}, including the tilt term, $\mathcal{T}$ in our mass model,  Eq.~(\ref{eqn:1dapprox}) becomes:

\begin{equation}
    \overline{v_z^2}(r) = \frac{1}{\nu(z)}\int_0^z \nu(z') \left(K_z(z') - \mathcal{T}(z')\right) dz' + \frac{C}{\nu(z)}
    \label{eqn:1dapproxtilt}
\end{equation}
As in \citet{Sivertsson:2017rkp}, we assume that the tracer population falls off exponentially with radius: $\nu(R,z) = \nu(z)\exp(-R/h)$ such that the tilt term can be simplified as:

\begin{equation} \label{tilt_eq}
\mathcal{T} = \frac{1}{R\nu}\frac{\partial}{\partial R}\left( R\nu\mathrm{ \ }\overline{v_Rv_z} \right) = \frac{\partial \left(\overline{v_Rv_z}\right)}{\partial R} + 
\left(\frac{1}{R} - \frac{1}{h} \right)\overline{v_Rv_z},
\end{equation}
where $h$ is the radial scale length of the tracer population. As we will see, it is not entirely correct to model $\nu(R)$ as a single exponential function for such a large range in $z$. However, the contribution of the tilt term turns out to be sufficiently small that this approximation will suffice.

In \citet{Silverwood:2015hxa} and \citet{Sivertsson:2017rkp}, we assumed a simple power law for the cross term of the velocity ellipsoid, $\overline{v_R v_z} = A\left(\frac{z}{\rm kpc}\right)^n$. As we shall show here, however, this is not sufficient to capture the behaviour of $\overline{v_R v_z}$ in the simulations. For this reason, in this paper we generalise the treatment of $\overline{v_R v_z}$, measuring it, along with the radial scale length, $h$, directly from the simulation data for each wedge. To integrate over $\mathcal{T}$ in Eq. (\ref{eqn:1dapproxtilt}), we fit a 3:rd order polynomial in $z$ to the simulation data, performing these fits separately for $z>0$ and $z<0$. This fit is performed jointly alongside the \MultiNest\ fit to the other model parameters in our Jeans analysis (\S\ref{sec:results_tilt}). As such, we show the tilt terms extracted from the simulations in this way later on the paper in Fig.~\ref{tilt_sgr600_rotc_tilt_fig}.

\subsection{The axial tilt term}\label{sec:axialtilt}
The axial tilt term, $\mathcal{A}$, can be treated similarly to the tilt term \citep{Silverwood:2015hxa}. However, this will only be worthwhile if it contributes significantly to the Jeans equations. To assess this, from Eq. (\ref{eqn:Kz}), we can cast $\mathcal{A}$ as an effective surface density term, $\Delta \Sigma_\mathcal{A}(R,z)$:

\begin{equation}
\mathcal{A} = \frac{1}{R\nu}\frac{\partial}{\partial \phi}\left(\nu\mathrm{ \ }\overline{v_\phi v_z} \right) = 2\pi G \Delta \Sigma_\mathcal{A}(R,z)
\label{eqn:axtiltsurf}
\end{equation}
Since $\mathcal{A}$ depends on the change of $\nu$ and $\overline{v_\phi v_z}$ with azimuth, $\phi$, $\Delta \Sigma_\mathcal{A}$ vanishes if the galaxy is perfectly axisymmetric. 

\begin{figure}
\includegraphics[width=\columnwidth, trim={0 0 46cm 0},clip]{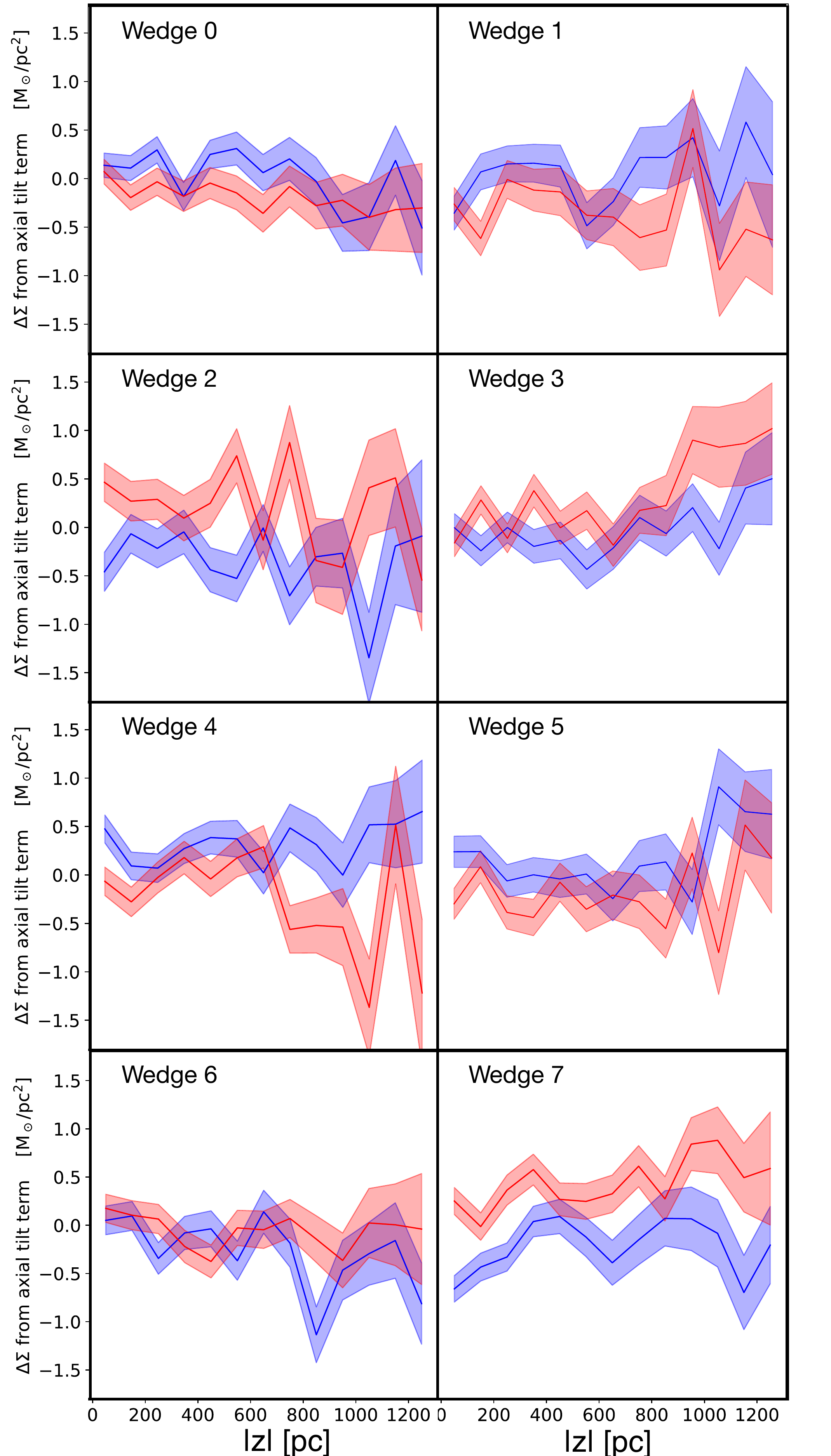}
\vspace{-4mm}
\caption{The size of the axial tilt term when interpreted as an effective surface density contribution, $\Delta \Sigma_\mathcal{A}$ (Eq. \ref{eqn:axtiltsurf}) for the \MWB\ simulation. The data points for the $z>0$ and $z<0$ directions are shown with black dots and stars, respectively. Note that the red and blue bands, for the $z>0$ and $z<0$ directions, respectively, do not correspond to any analysis or fit; they are just to guide the eye. As compared to the total surface density (Fig.~\ref{Sgr600_first_surfdens}), $\Delta \Sigma_\mathcal{A}$ is very small (of order 1\% of the total) and so the axial tilt term can be safely neglected in our analysis.}
\vspace{-4mm}
\label{Sgr600_axial_tilt}
\end{figure}

In Fig.~\ref{Sgr600_axial_tilt}, we show $\Delta \Sigma_\mathcal{A}$ for the \MWB\ simulation. This shows that the contribution of the axial tilt term is very small as compared to the true surface densities (Figs.~\ref{GSE_first_surfdens} and \ref{Sgr600_first_surfdens}). (The results for the \MWA\ simulation are similar and so we omit these for brevity.) We conclude that the contribution from the axial tilt term can safely be neglected.

\subsection{The explicit time dependent term}\label{sec:timedep}

The explicit time dependent term in the Jeans Eq.~(\ref{Jeans_eq}) is challenging to measure from observations of a real galaxy, though it can be estimated by a careful elimination of all other terms, or through modelling inconsistencies between tracer populations with distinct vertical frequencies \citep{2017MNRAS.464.3775B}. In this paper, we take the former approach. By including all terms in the Jeans equations, any remaining bias in our recovery of $\rhodm$ must owe to the time dependent term.

\subsection{The tracer and mass density} \label{Mass_dens_modelling}
 
Most studies in the literature to date assume an exponential form for the tracer density profile $\nu \propto \exp(-|z|/z_0)$ which greatly simplifies the Jeans equations \citep[e.g.][]{Haines2019}. However, we find that our perturbed $N$-body simulations can often depart significantly from this simple assumption. As such, we assume a much more general form for $\nu$ by fitting the $\nu$ data with several exponential segments, each over a smaller range in $z$. Each segment in the $\nu$ fit is given by an exponential function with its own scale height and normalization, adjusted so that the fit is continuous. The number of segments is adjusted so that a good fit to the tracer density is achievable while still maintaining a smooth function. This more general treatment of $\nu$ is possible thanks to our choice of solving the combination the Jeans-Poisson equations by integration rather than by differentiation (Eq.~\ref{eqn:1dapprox}), making it straightforward for us to integrate the segment-wise fit to $\nu$. It also has the  advantage that the integrals in Eqs. (\ref{eqn:1dapprox}) and \ref{eqn:Kz} are analytic. An example segment-wise fit to the tracer density data is shown in Fig.~\ref{fig:nu-dens-rotc-tilt}.

We divide the total matter density, $\rho$, into a contribution from baryons, $\rhob$, and dark matter, $\rhodm$: $\rho=\rhob + \rhodm$. We assume that $\rhodm$ is constant in each wedge\footnote{In reality, one expects $\rhodm$ to decrease slowly with $z$, since a larger $z$ also implies a larger distance to the galactic centre. However, for the common assumption that the galactic dark matter halo follows an NFW profile \cite{Navarro:1996gj}, the $\rhodm$ dependence on $z$ is negligible for the $z$ values that we consider \citep[e.g.][]{Garbari2011}.}, and we assume that $\rho_b$ has the same shape as $\nu$, with an uncertainty of 20\% on its normalisation. Note that under these definitions, $\rhodm$ will include any contribution to the potential from stars and dark matter accreted from the satellite merger.

\subsection{Model fitting, likelihood function and priors}\label{sec:liklihood}

We scan through the parameter space using the Bayesian nested sampling tool \MultiNest\ \citep{multinest2019, pymultinest2014} to find which model parameters provide the best fit to the observables $\vzsqmean$ and $\nu$, as well as $\mathcal{T}$, when applicable.

The natural logarithm of our likelihood is defined as:
\begin{equation}\label{eq:lnlike}
\ln \mathcal{L}
= -\frac{1}{2} \left( \chi^2(\nu) + \chi^2 (\overline{v_z^2}) + \chi^2(\mathcal{T}) 
\right),
\end{equation}
with:
\begin{equation}
    \chi^2(A) = \sum_i \frac{\left( A_{\mathrm{data},i} - A_{\mathrm{model},i} \right)^2}{\sigma^2_{A,i}},
\end{equation}
where $i$ iterates over all data bins (binned by $z$), and $\sigma_{A,i}$ corresponds to the standard deviation of the quantity $A$ in $i^{th}$ bin (that in this paper owes just to sampling error). The first two terms of Eq.~\eqref{eq:lnlike} will be taken into account in all of our analyses. The third term, $\chi^2(\mathcal{T})$, will be included when we additionally model the tilt term of the vertical Jeans Eq.~\eqref{Jeans_eq}.

We include $\overline{v_z^2}$ data up to approximately 2 scale heights in $\nu$ and $\rho_b$, in line with \cite{Garbari2011}. The scale heights vary significantly between the $N$-body simulations investigated here, in part because perturbations puff up the disc and hence increase the scale height. The individual limit of 2 scale heights maximizes the use of the data while avoiding the highest $|z|$ bins that have very few stars 
and hence large statistical noise and bias. 
For $\nu$, however, we include slightly larger values of $|z|$ because the upper limit of the integral in Eq.~\eqref{simple_Jeans_Poisson} goes to infinity. Even though the contribution from higher $|z|$ values is small because of the exponential fall-off of $\nu$, the values of $\nu$ from the $z$ regions just outside of the range used in $v_z^2$ data can still have some impact on the evaluation of $v_z^2$  from Eq.~(\ref{simple_Jeans_Poisson}).

We set a linear prior on the dark matter density in the range $10^{-3} < \rhodm / ({\rm M}_\odot\,{\rm pc}^{-3}) < 10^{-1}$. We use a 20\% flat prior on the baryon density around the true value. For the tracer density priors, we make an initial fit to the data, setting 50\% flat priors on the best-fit scale heights and a logarithmic prior on the normalisation of $\pm 2.5$\,dex around the true value. For the tilt, we make a polynomial fit to the data and then set a 50\% flat prior around the best fit values. For the rotation curve term, we fix the values to those extracted from the simulations -- a best-case scenario.

\section{Results}\label{sec:results}

In this section, we show the results from applying our Jeans method to the \MWA\ and \MWB\ simulations, with and without the inclusion of the rotation curve and tilt terms in the steady-state Jeans-Poisson Eqs. (\ref{Jeans_eq} and \ref{Poisson_eq}). (As discussed in \S\ref{sec:jeans}, the 2:nd Poisson term and the axial tilt term are small and so we neglect these.)

\subsection{The 1D approximation}\label{sec:results_1Dapprox}

\begin{figure}
\centering
\includegraphics[width=\columnwidth]{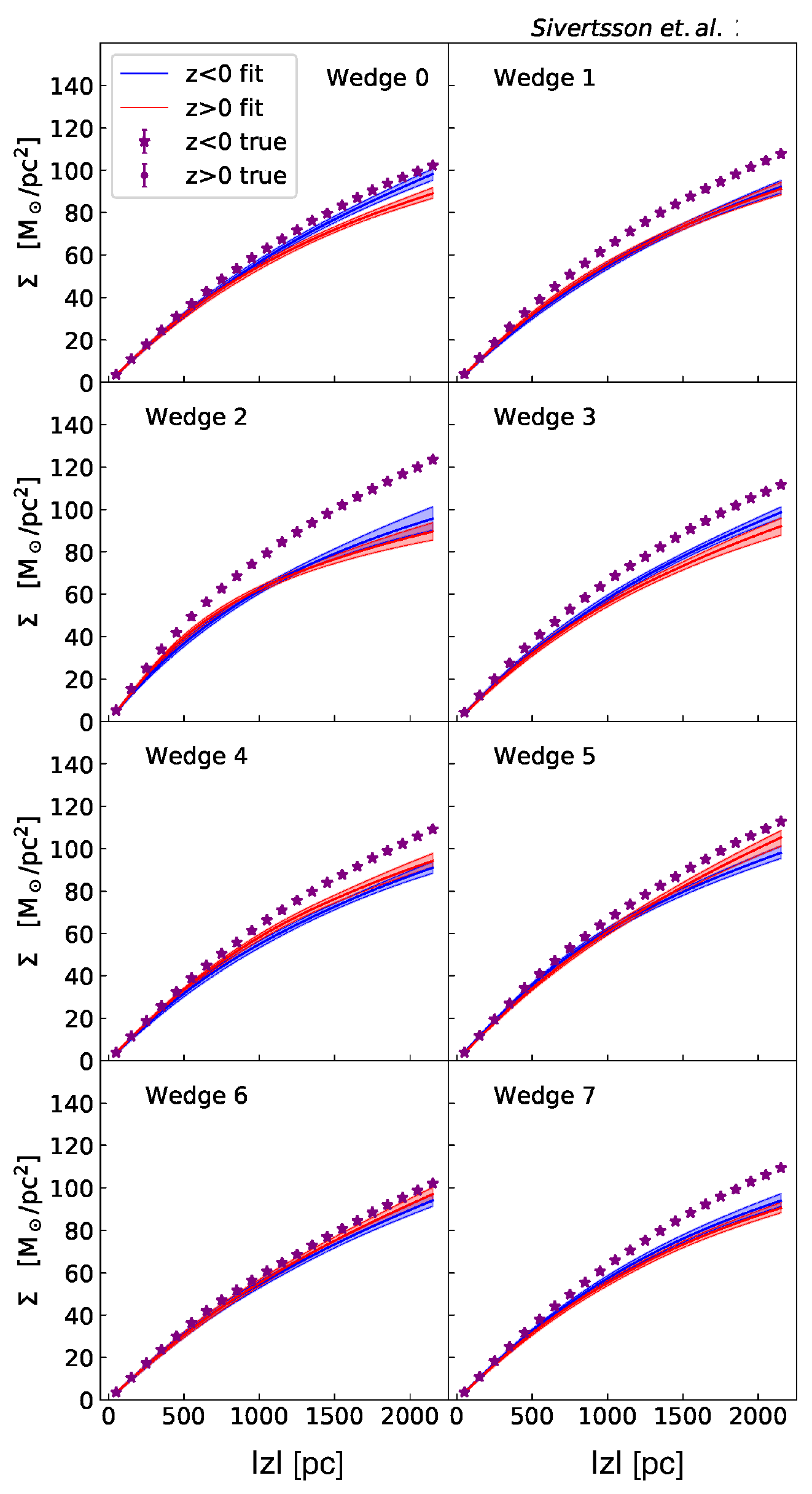}
\vspace{-4mm}
\caption{Recovered surface densities from Jeans mass modelling the \MWA\ simulation. In this plot, we use the `1D approximation' that neglects the tilt and rotation curve contributions to the Jeans-Poisson Eqs. (Eqs.~\ref{simple_Jeans_Poisson} and \S\ref{sec:1dapprox}). Each panel shows one of the 8 wedges shown in Fig.~\ref{stellar_locations}. The blue and red bands show the 95\% credible regions for the recovered surface densities from the \MultiNest\ runs in the $z<0$ and $z>0$ regions, respectively. The true values are shown as purple stars and dots, 
again for $z<0$ and $z>0$ regions, respectively (the stars and dots in this plot overlap to the point where they are indistinguishable).}
\vspace{-4mm}
\label{GSE_first_surfdens}
\end{figure}

\begin{figure}
\centering
\includegraphics[width=\columnwidth]{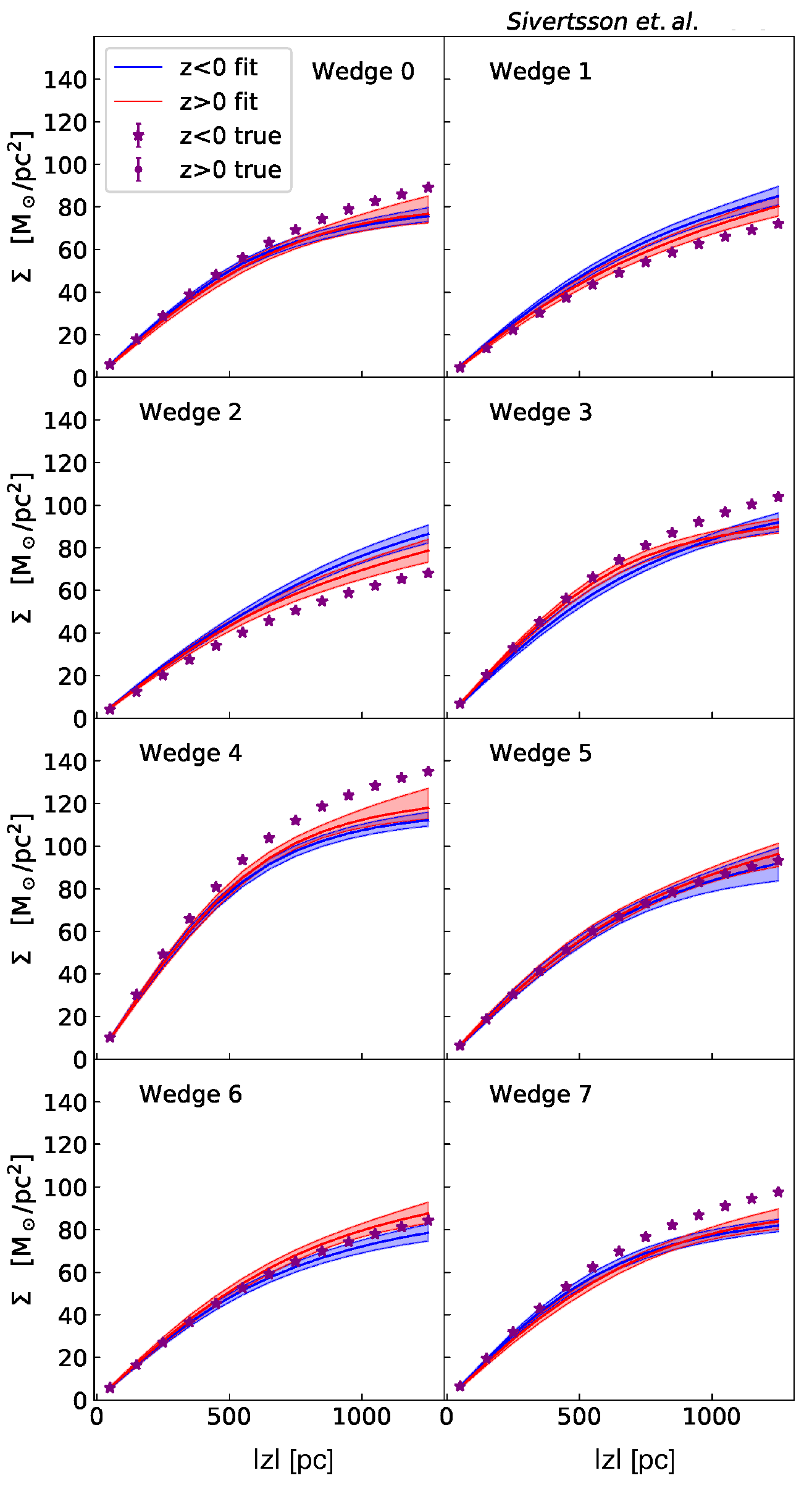} 
\vspace{-4mm}  
\caption{Recovered surface densities from Jeans mass modelling the \MWB\ simulation in the `1D approximation'. Lines and points are as in Figure \ref{GSE_first_surfdens}.}
\vspace{-4mm}
\label{Sgr600_first_surfdens}
\end{figure}

We begin by considering first the `1D approximation' typically employed in the literature to date (\S\ref{sec:1dapprox}). In Figs.~\ref{GSE_first_surfdens} and \ref{Sgr600_first_surfdens}, we show the recovered surface  densities\footnote{Note that `surface density' here refers to $2\int_0^z\rho(z')\ud z'$ and $2\int_{-z}^0\rho(z')\ud z'$ for the $z>0$ and $z<0$ directions, respectively, and not to the more usual
definition $\Sigma(z) = \int_{-z}^z\rho(z')\ud z'$.\label{surfdens_footnote}} for the \MWA\ and \MWB\ simulations, respectively. As can be seen, for the \MWA\ simulation, the results are generally rather poor. In all wedges, the surface densities are underestimated at more than 95\% confidence and in some cases by up to a factor of $\sim 1.5$ (wedge 2). Only wedges 0 and 6 recover the true surface density within the 95\% confidence interval, and then only for $z<0$ and $z > 0$, respectively. Worse, with the exception of wedge 0, the confidence intervals overlap for the $z<0$ and $z>0$ samples. Discrepancies between the dynamics of $z<0$ and $z>0$ stars has been invoked in the literature as a mark of disequilibrium dynamics \citep[e.g.][]{Haines2019,Salomon20} and so the lack of such a signature could yield false confidence in the results.

The situation is similar for the \MWB\ simulation (Fig.~\ref{Sgr600_first_surfdens}). Only wedges 5 and 6 recover the surface density within the 95\% confidence band, while the worst wedge (4) is biased low by a factor of $\sim 1.4$ for the $z<0$ tracers.

\subsection{Including the rotation curve term}\label{sec:results_rotcurve}

\begin{figure}
\includegraphics[width=\columnwidth]
{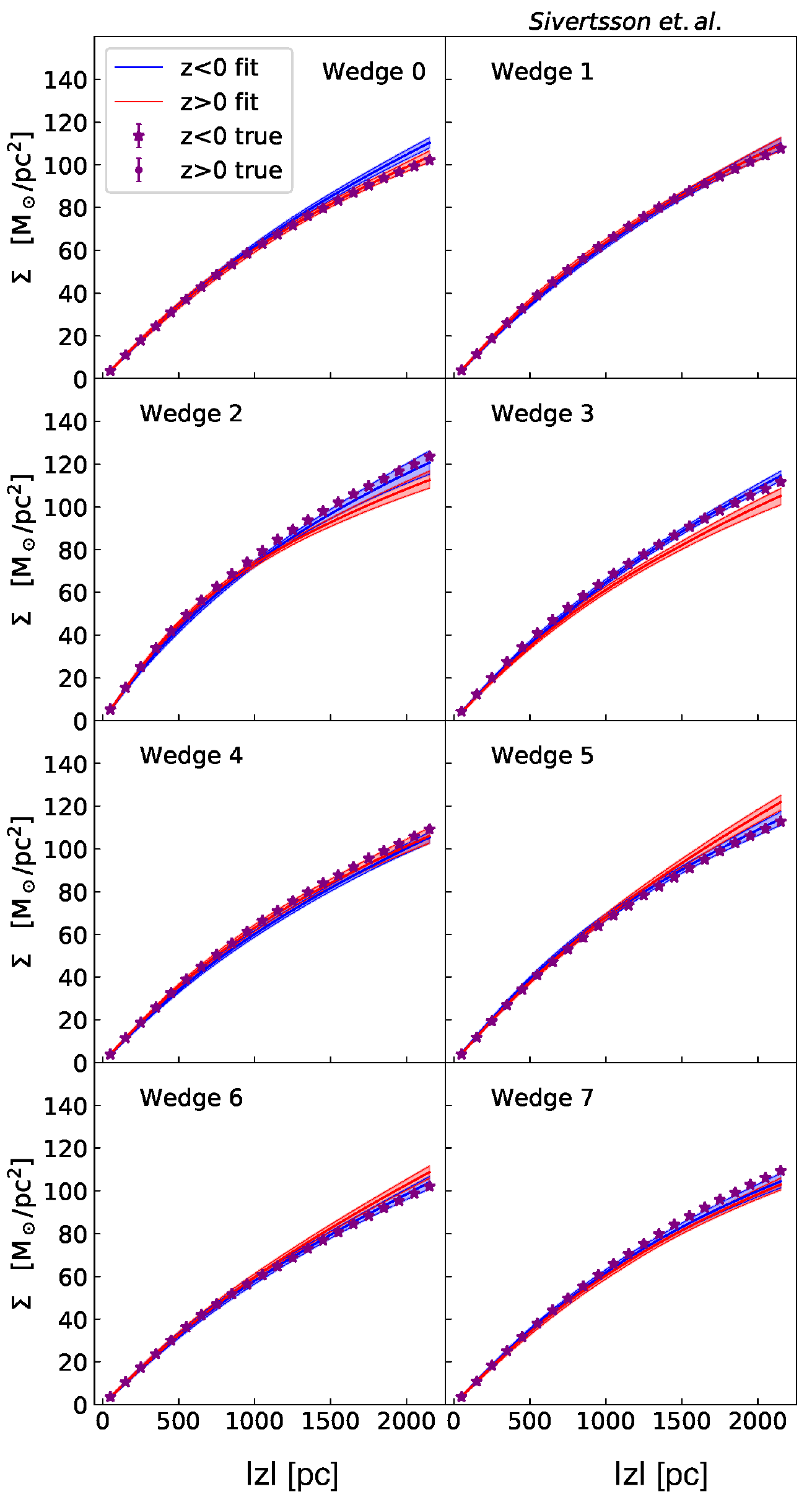}
\vspace{-4mm}
\caption{Recovered surface densities from Jeans mass modelling the \MWA\ simulation in the `1D approximation', including the rotation curve term. Lines and points are as in Figure \ref{GSE_first_surfdens}. Notice that the recovery is now much improved as compared to the analysis without the rotation curve term (Fig.~\ref{GSE_first_surfdens}).}
\vspace{-4mm}
\label{Sigma_GSE_rotc_fig}
\end{figure}

\begin{figure}
\includegraphics[width=\columnwidth]
{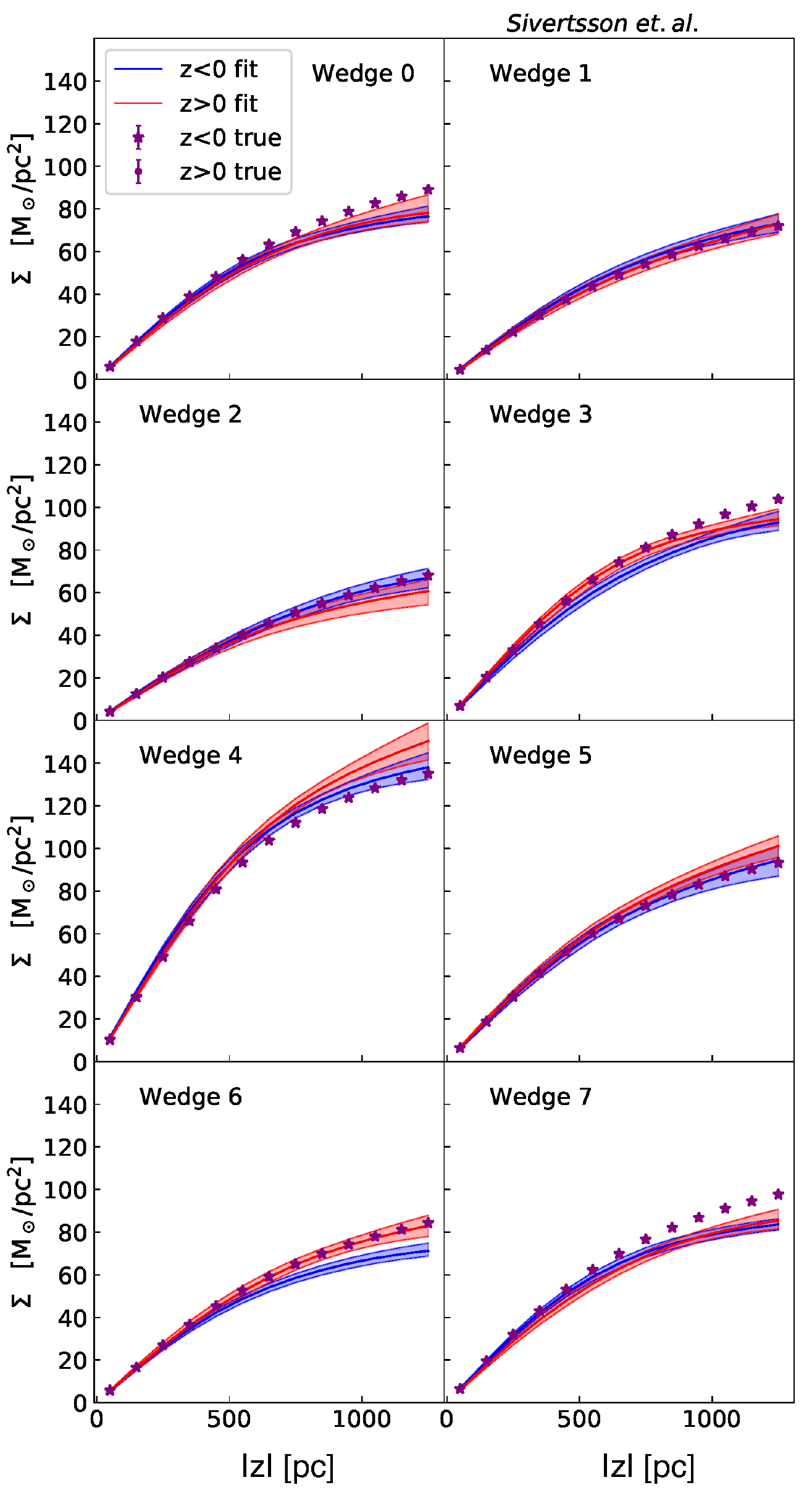}
\vspace{-4mm}
\caption{Recovered surface densities from Jeans mass modelling the \MWB\ simulation in the `1D approximation', including the rotation curve term. Lines and points are as in Figure \ref{GSE_first_surfdens}. Notice that the recovery is now much improved as compared to the analysis without the rotation curve term (Fig.~\ref{Sgr600_first_surfdens}).}
\vspace{-4mm}
\label{Sgr600_with_rotcurve}
\end{figure}

In 
Figs.~\ref{Sigma_GSE_rotc_fig} and \ref{Sgr600_with_rotcurve}, we show the recovery of the surface density when including the rotation curve term (Eq. \ref{eqn:rotcurveterm} and \S\ref{sec:rotcurve}). Notice that the recovery is now much improved for all wedges across both simulations, demonstrating the importance of including this term in the models. For \MWA, we now recover the input solution within the 95\% confidence bands for all wedges, except 2 and 3 for $z>0$. Even in these cases, the \pfds{statistical} bias is much smaller than previously, of the order 5\%. 
There is a similarly good improvement for simulation \MWB, though wedges 3, 6 and 7 all retain a bias larger than the 95\% intervals.

\subsection{Including the tilt term}\label{sec:results_tilt}

\begin{figure}
\includegraphics[width=\columnwidth]
{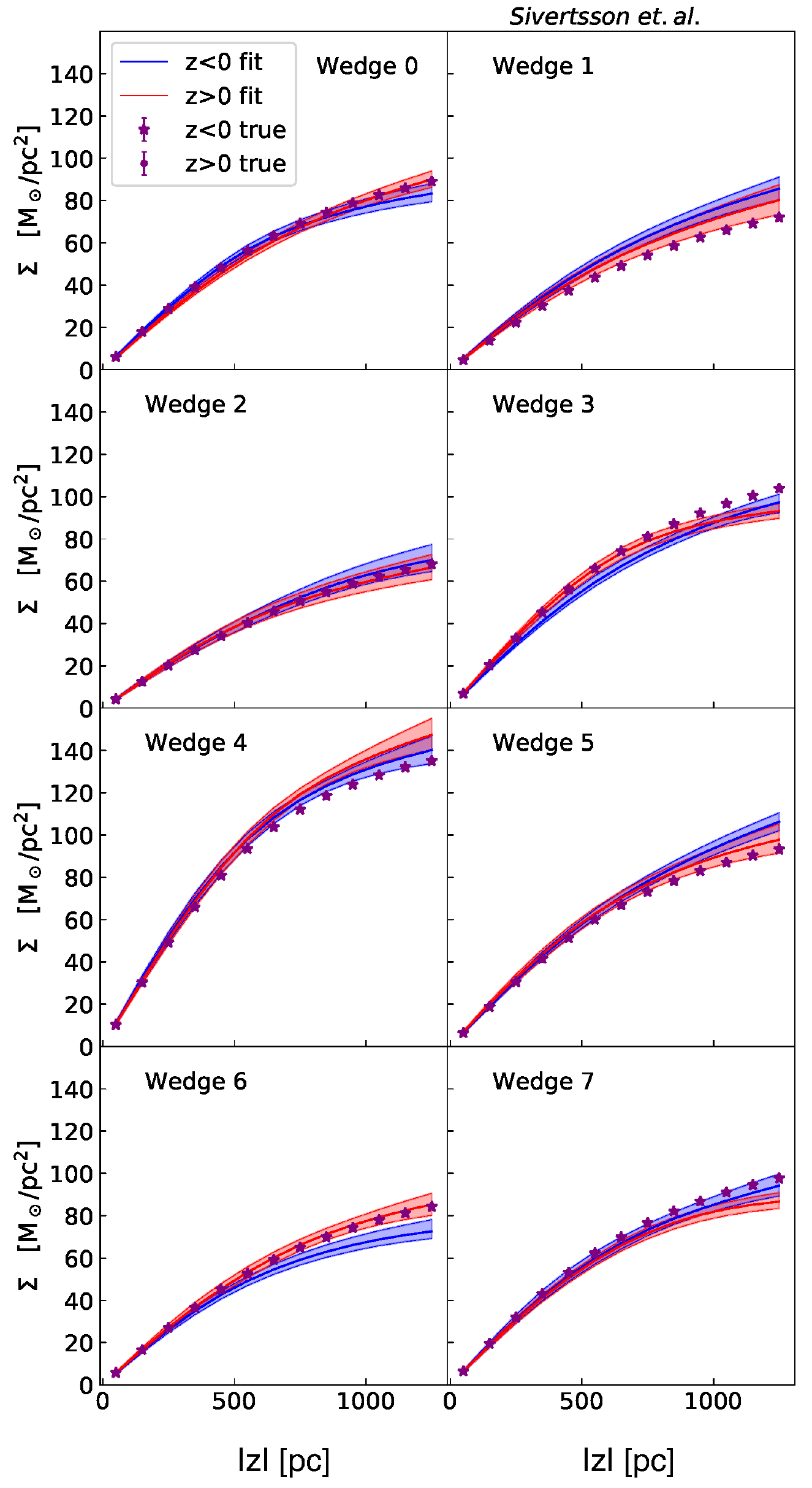}
\vspace{-4mm}
\caption{Recovered surface densities from Jeans mass modelling the \MWB\ simulation in the `1D approximation', including the rotation curve and tilt terms. Lines and points are as in Figure \ref{GSE_first_surfdens}. Notice that the recovery is now further improved as compared to the analysis with just the rotation curve term (Fig.~\ref{Sgr600_with_rotcurve}).}
\vspace{-4mm}
\label{Sigma_sgr600_rotc_tilt_fig}
\end{figure}

Fig.~\ref{Sigma_sgr600_rotc_tilt_fig} shows our recovery of the surface density when including the rotation curve and tilt terms in our analysis (Eq.~\ref{eqn:1dapproxtilt} and \S\ref{sec:tilt}). This further improves our recovery of the input solution, demonstrating the importance also of this term. Now, combining the scatter in solutions from the $z<0$ and $z>0$ samples, there is no longer any bias larger than our 95\% confidence bands. There can still be bias, however, if only one of these samples is considered. For example, the $z<0$ tracers in wedge 6 are systematically biased low. (We obtain already unbiased results for the \MWA\ simulation without the tilt term (Fig.~\ref{Sigma_GSE_rotc_fig}), and so omit an analysis of the \MWA\ simulation including tilt, for brevity.)

\begin{figure}
\includegraphics[width=\columnwidth]
{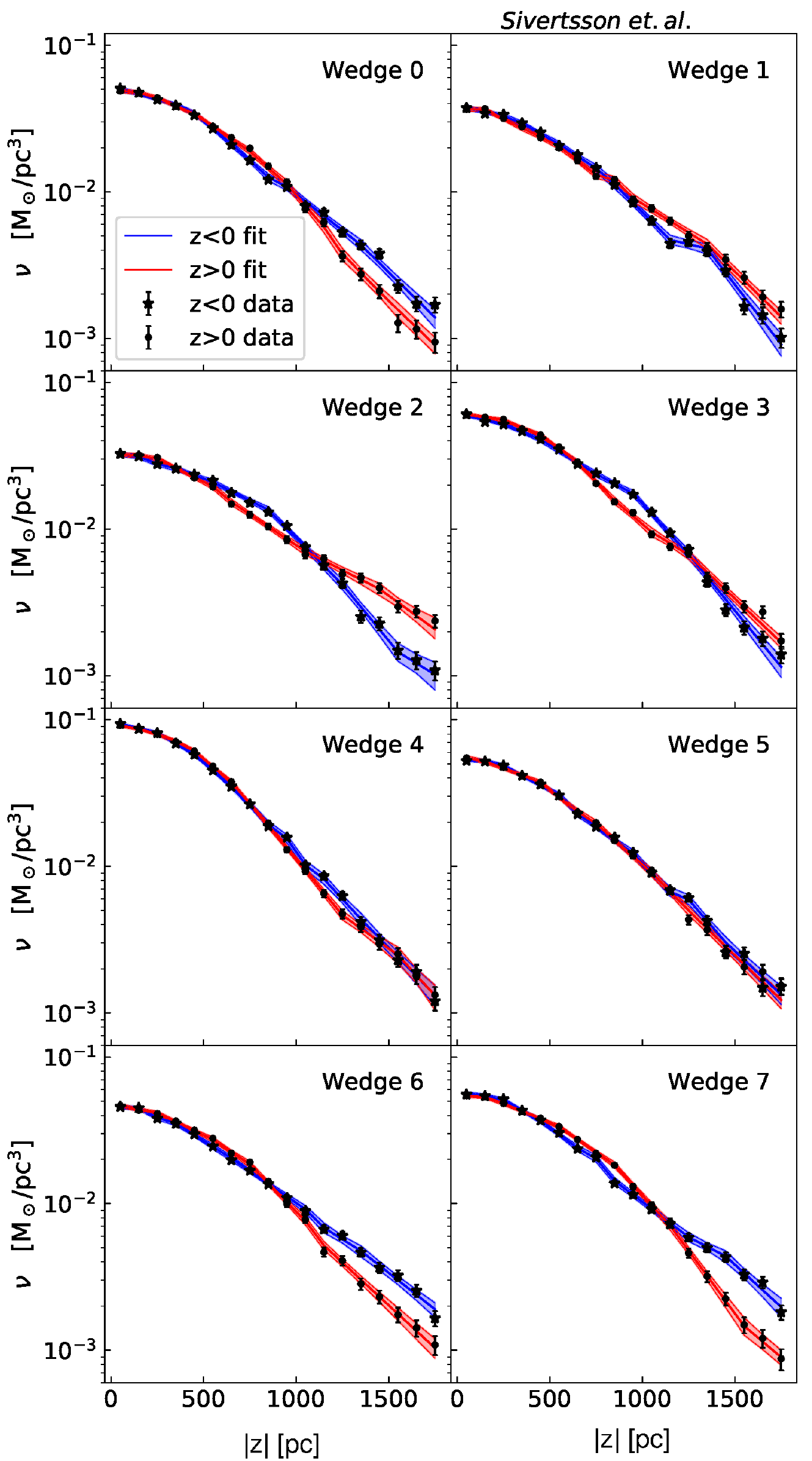}
\vspace{-4mm}
\caption{Example fit to the tracer densities from the same analysis of the \MWB\ simulation as shown in Fig.~\ref{Sigma_sgr600_rotc_tilt_fig} (i.e. when including both the rotation curve and the tilt terms in the analysis). Conventions are as in Figs.~\ref{Sgr600_first_surfdens} and \ref{sigz_sgr600_rotc_tilt_fig}, but showing $\nu$ instead of $\Sigma$. Note that the fit to the $\nu$ data includes higher $|z|$ values than the fit to the $\overline{v_z^2}$ data, to increase the accuracy of our method (see \S\ref{sec:liklihood} for details). The error bars on the data points include only Poisson noise and represent 1 standard deviation.}
\vspace{-4mm}
\label{fig:nu-dens-rotc-tilt}
\end{figure}

\begin{figure}
\includegraphics[width=\columnwidth]
{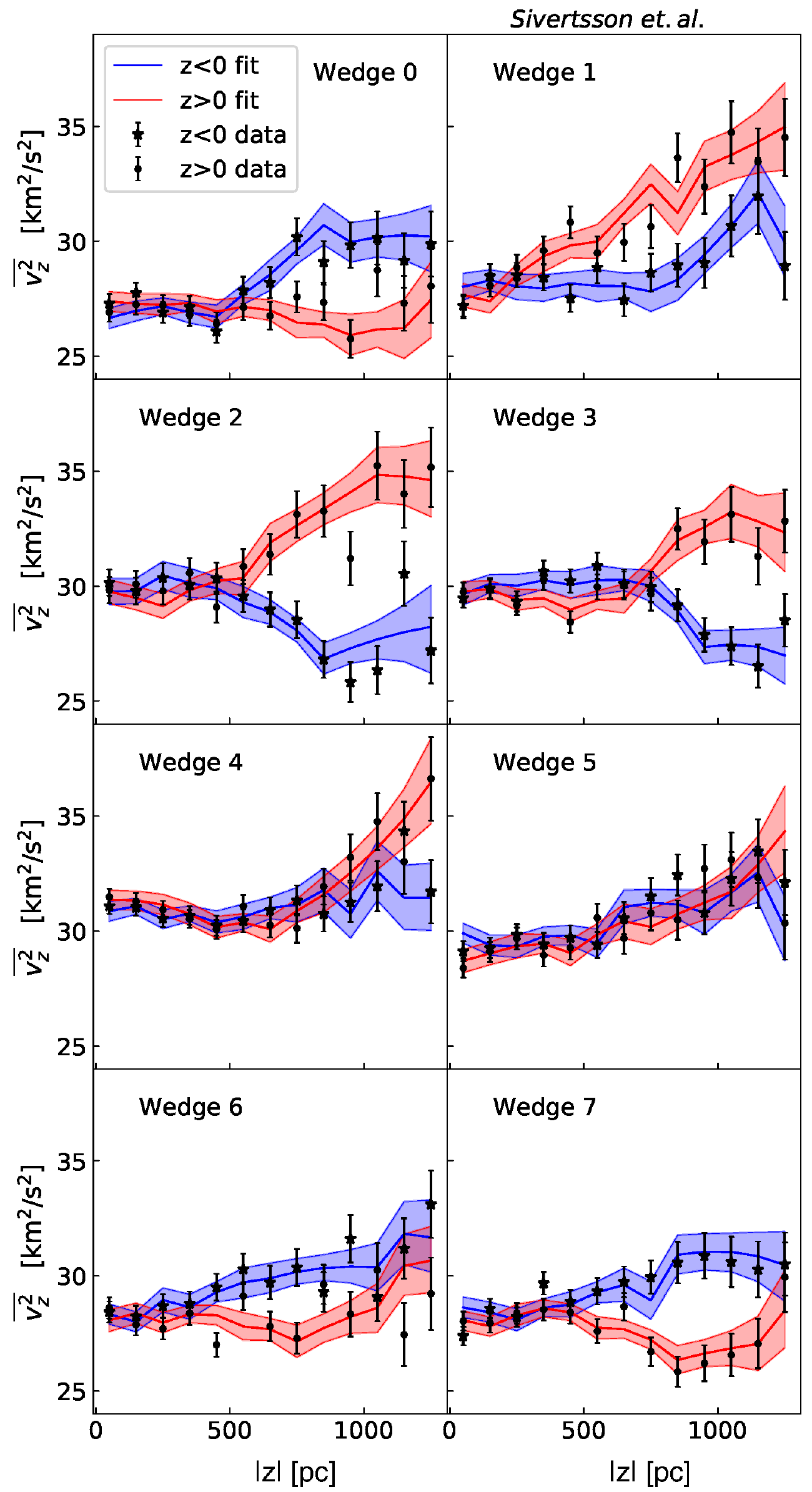}
\vspace{-4mm}
\caption{Example fit to the vertical velocity dispersions from the same analysis of the \MWB\ simulation as shown in Fig.~\ref{Sigma_sgr600_rotc_tilt_fig} (i.e. when including both the rotation curve and tilt terms in the analysis). Conventions are as in Fig.~\ref{Sgr600_first_surfdens} but showing $\overline{v_z^2}$ instead of $\Sigma$. The error bars on the data points include only Poisson noise and represent 1 standard deviation.}
\vspace{-4mm}
\label{sigz_sgr600_rotc_tilt_fig}
\end{figure}

\begin{figure}
\includegraphics[width=\columnwidth]
{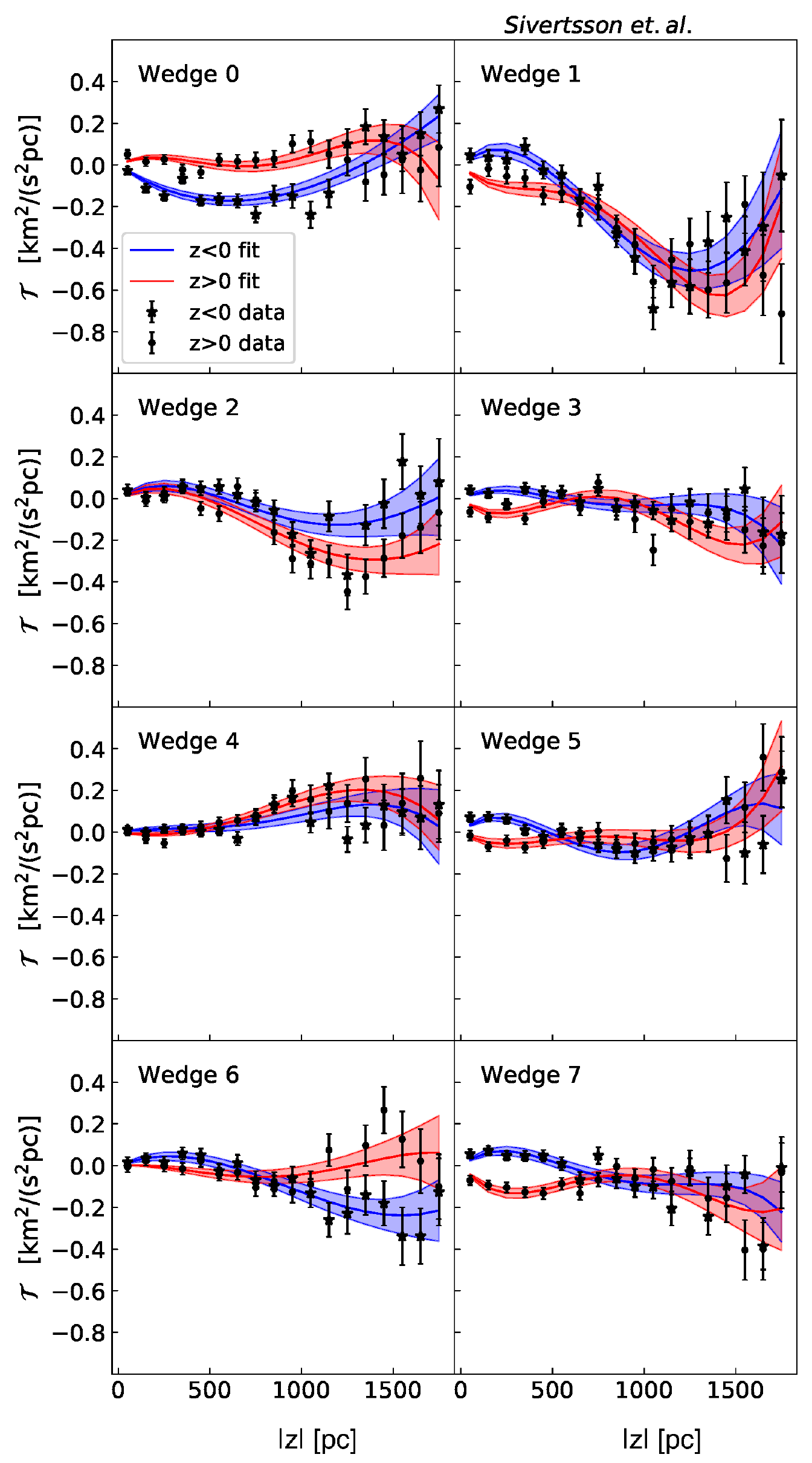}
\vspace{-4mm}
\caption{Example fit to the tilt term from the same analysis of the \MWB\ simulation as shown in Fig.~\ref{Sigma_sgr600_rotc_tilt_fig} (i.e. when including the rotation curve and tilt terms). Conventions are as in Figs.~\ref{Sgr600_first_surfdens} and \ref{sigz_sgr600_rotc_tilt_fig}, but showing $\mathcal{T}$ instead of $\Sigma$. The tilt terms are modelled as 3:rd order polynomials. As for $\nu$, we include higher $|z|$ data points in the $\mathcal{T}$ fit (see \S\ref{sec:liklihood} for details). The error bars on the data points include only Poisson noise and represent 1 standard deviation.}
\vspace{-4mm}
\label{tilt_sgr600_rotc_tilt_fig}
\end{figure}

\begin{figure}
\includegraphics[width=\columnwidth, trim={0.15cm 0.0cm 0.8cm 0.05cm},clip]{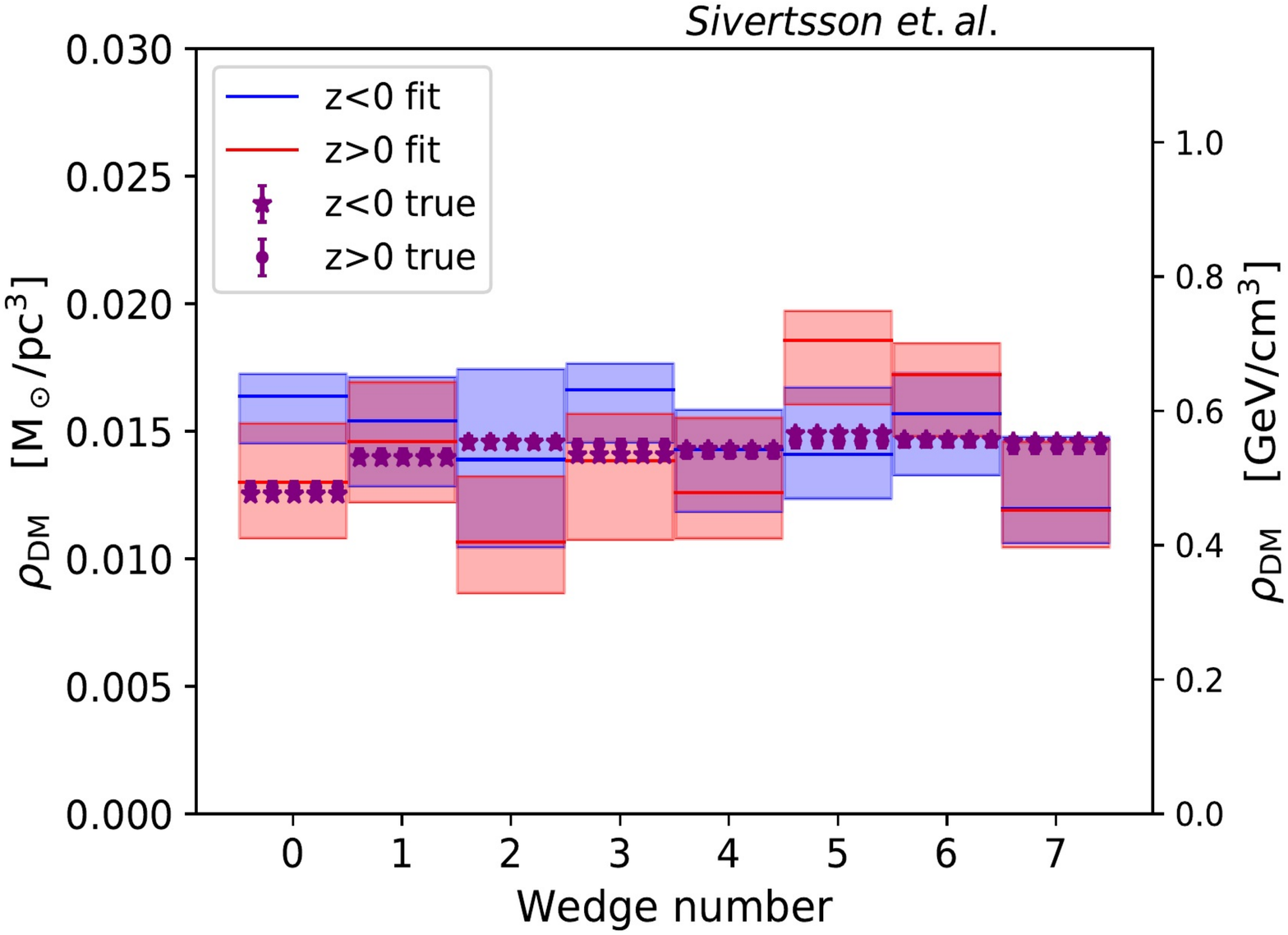}
\includegraphics[width=\columnwidth, trim={0.15cm 0.0cm 0.8cm 0.05cm},clip]{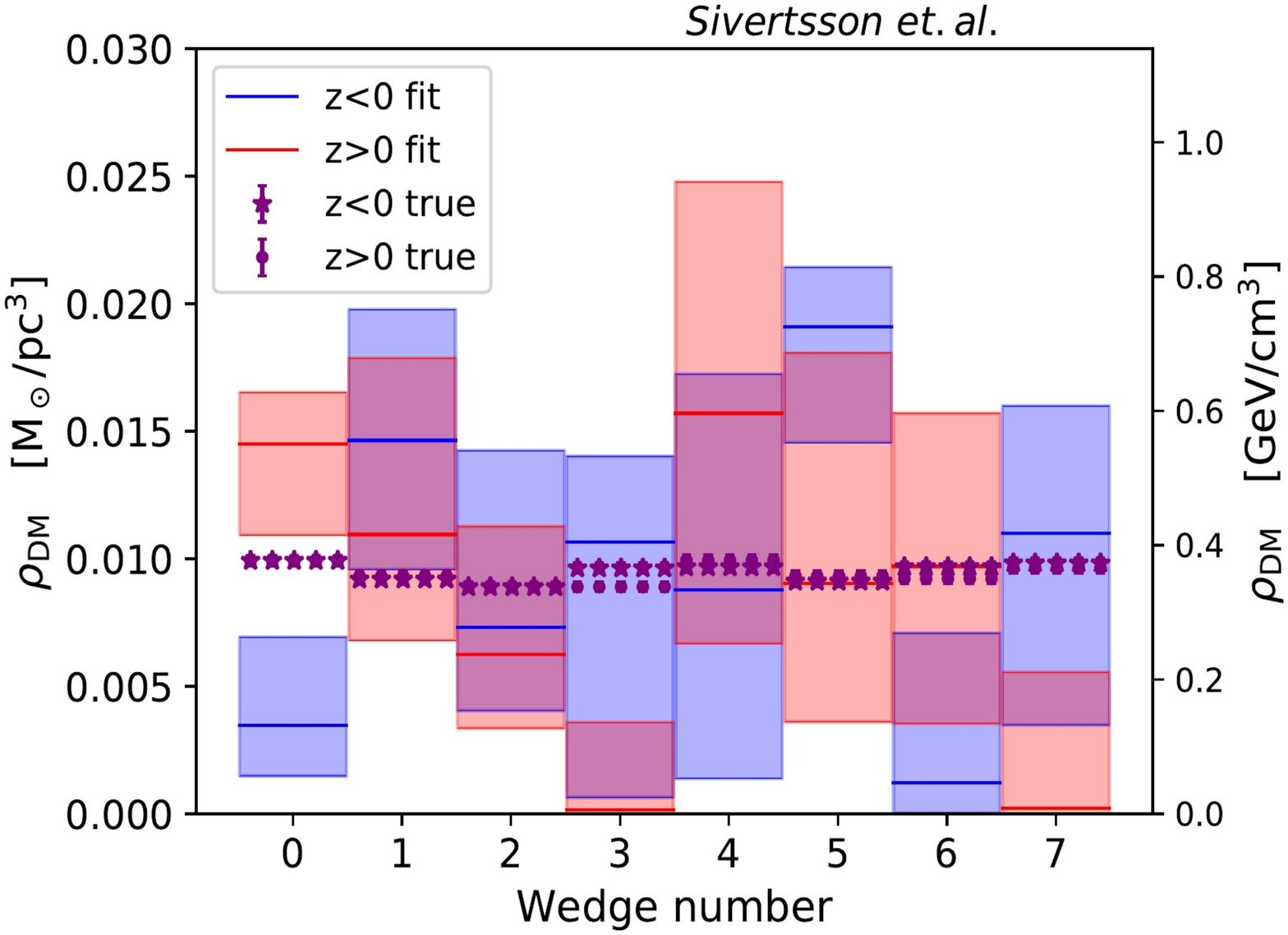}
\includegraphics[width=\columnwidth, trim={0.15cm 0.0cm 0.8cm 0.05cm},clip]{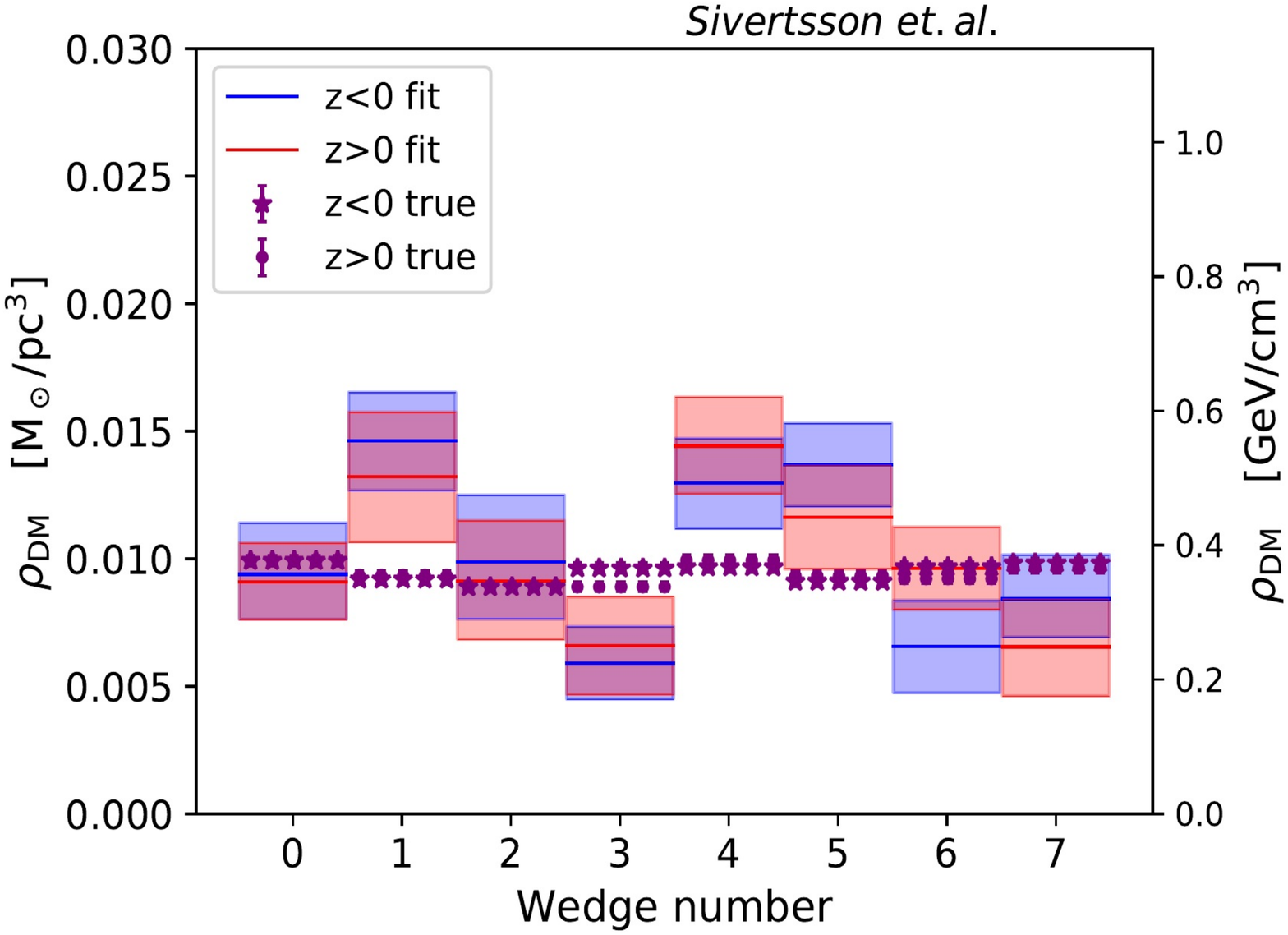}
\vspace{-4mm}
\caption{Recovery of the local dark matter density from Jeans mass modelling the \MWA\ (upper, including the rotation curve term), \MWB\ (middle, including the rotation curve and tilt terms) and \MWB\ (bottom, as the middle panel but using a tighter 1\% prior on the baryon density) simulations, shown as a function of wedge number. The red and blue bands show the recovered values of $\rhodm$ with 2$\sigma$ error bars for the $z>0$ and $z<0$ samples, respectively, while the purple dots and stars show the true values (from the simulations).  The top two panels
assume an error on baryonic surface density $\Sigma_b$ of 20\%, comparable to current estimates; the bottom panel
is relevant for the future once the baryon density is known to 1\%.  Notice that the uncertainties are smaller for the \MWA\ simulation due to the increased sampling in each wedge, while there is more bias in the \MWB\ simulation that becomes more visible once tighter priors are placed on the baryon density (bottom panel).}
\vspace{-4mm}
\label{rhoDM_Sgr600_inc_rotterm_tilt}
\end{figure}

\begin{figure}
\includegraphics[width=\columnwidth]{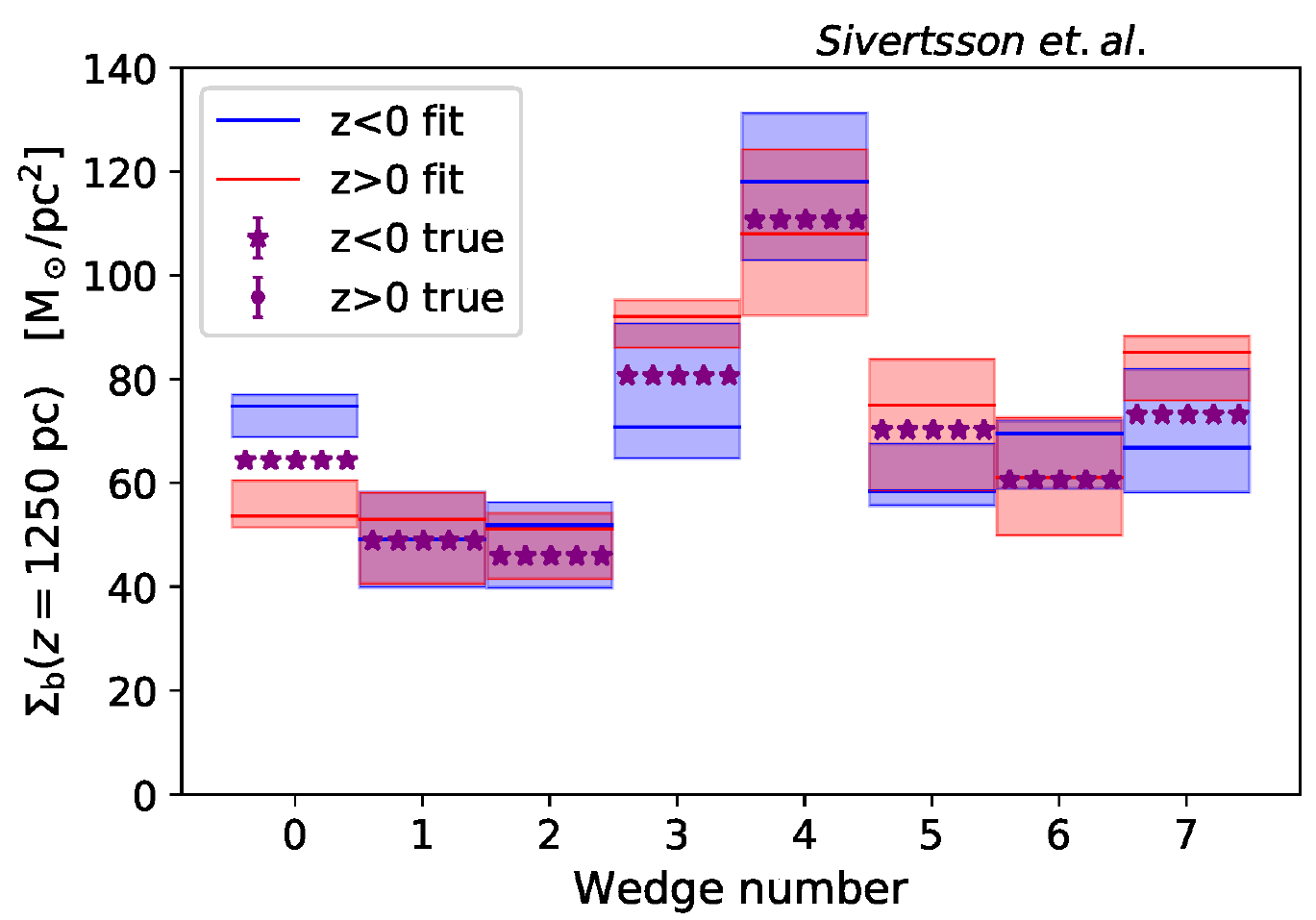}
\vspace{-4mm}
\caption{Recovery of the baryonic surface density at $z=\pm 1250$\,pc (defined as $2\int_{-1250}^0\rho\ud z$ (blue bands) and $2\int_0^{1250}\rho\ud z$ (red bands)), for the same analyses of the \MWB\ simulation as shown in Figs.~\ref{Sigma_sgr600_rotc_tilt_fig} and \ref{rhoDM_Sgr600_inc_rotterm_tilt}  (i.e. including both the rotation curve and tilt terms). The purple dots and stars show the true values for the $z>0$ and $z<0$ regions, respectively. 
Notice that for wedge 0 in which $\Sigma_b$ is offset from the true value (too high for $z<0$ and too low for $z>0$), the direction of the offset anti-correlates with that of the dark matter density shown in Fig.~\ref{rhoDM_Sgr600_inc_rotterm_tilt}, indicating that the total surface mass density (sum of baryons and dark matter) has been correctly recovered for this wedge.}
\vspace{-4mm}
\label{Sigma_bary_Sgr600_inc_rotterm_tilt}
\end{figure}

The quality of our fits to the tracer density $\nu(z)$, vertical velocity dispersion $\overline{v_z^2}(r)$, and tilt term for the \MWB\ simulation are demonstrated in Figs.~\ref{fig:nu-dens-rotc-tilt}, \ref{sigz_sgr600_rotc_tilt_fig} and \ref{tilt_sgr600_rotc_tilt_fig}, showing that this model gives an excellent recovery of the data for all wedges. In particular, notice that there is no obvious feature in the data that can predict which wedges will be most biased. Wedge 6 for the $z<0$ tracers, for example, remains most biased after the inclusion of the rotation curve and tilt terms, but shows no particular feature in the data as compared to the other wedges.

\subsection{Recovering the local dark matter density}\label{sec:localdm}

In Figs~\ref{rhoDM_Sgr600_inc_rotterm_tilt}, we show our recovery of $\rhodm$ for \MWA\ (upper, including the rotation curve term), \MWB\ (middle, including the rotation curve and tilt terms) and \MWB\ (bottom, as the middle panel but using a tighter 1\% prior on the baryon density) simulations, shown as a function of wedge number.

Firstly, notice that with our default 20\% errors on $\Sigma_b$ (top and middle panels) we recover $\rhodm$ within our quoted uncertainties for almost all wedges in both simulations, if we consider the combined uncertainty from the $z>0$ and $z<0$ analysis. Only wedge 0 for the \MWB\ simulation shows statistically significant bias, being too high above the disc and too low below the disc. In Fig.~\ref{Sigma_bary_Sgr600_inc_rotterm_tilt}, we show the corresponding recovery of $\Sigma_b$ for the \MWB\ simulation. In this plot, we see the same effect for wedge 0 but anti-correlated with the recovery of $\rhodm$. This indicates that we have correctly recovered the {\it total} surface density, but partitioned it incorrectly into dark matter and baryons for this wedge. This highlights the known degeneracy between $\Sigma_b$ and $\rhodm$ \citep{Sivertsson:2017rkp}. We discuss this further in \S\ref{sec:discussion}.

While the bias on $\rhodm$ is acceptably small in both simulations, the 95\% confidence intervals for the \MWB\ simulation are, however, substantial, with uncertainties on $\rhodm$ of up to a factor of two. Three wedges are consistent with no detection of $\rhodm$ at 95\% confidence. Such large uncertainties owe to the sample error and the uncertainty on $\Sigma_b$. For \MWA, the wedges have $\sim 4$ times more stars than in \MWB. Assuming Poisson errors, this should yield an uncertainty on $\rhodm$ a factor $\sim 2$ smaller, consistent with what we see in Fig.~\ref{rhoDM_Sgr600_inc_rotterm_tilt} (compare the upper and middle panels). Reducing the uncertainty on $\Sigma_b$ to 1\% also reduces the uncertainties (Fig.~\ref{rhoDM_Sgr600_inc_rotterm_tilt}; bottom panel). Now, however, for simulation \MWB\ this exposes some statistically significant remaining bias for wedges 1 and 3 and, to a lesser extent wedge 4. We discuss this further, next.

\section{Discussion}
\label{sec:discussion}

\subsection{Comparison with previous work}

We have shown that our free-form Jeans analysis method, when including the tilt and rotation curve terms, recovers the surface mass density within our 95\% confidence intervals for a realistically perturbed Milky Way-like system, even in the face of significant deviations from axisymmetry and a `wobbling' disc (Fig.~\ref{Sigma_sgr600_rotc_tilt_fig}). At first sight this appears to be in contradiction with \citet{Haines2019}. They also apply a Jeans method to the \MWB\ simulation, finding instead that the surface density can be highly biased by up to a factor of $\sim 1.5$. However, the apparent difference between \citet{Haines2019} and our study owes to different assumptions used in our Jeans modelling. If we employ the `1D approximation' that neglects the tilt and rotation curve terms then we find a similar bias to \citet{Haines2019} in some of our wedges (Fig.~\ref{Sgr600_first_surfdens}). The \citet{Haines2019} analysis, based on \citet{Hagen_Helmi2018}, assumes that the tracer density distribution is exponential in $z$ and $R$, the radial and vertical velocity dispersion-squared are exponential in $R$ with the same scale length as the
density, and that the velocity ellipsoid tilt angle is constant with radius. As we have shown here, assumptions like these lead to much larger systematic errors than any non-steady state effect, at least in the \MWA\ and \MWB\ simulations.

\subsection{The rotation curve term}\label{sec:how-to-add-rot-curve}

We have shown that, for the \MWA\ and \MWB\ simulations that we analysed in this work, the `rotation curve' term, $\mathcal{R}$ (Eq.~\ref{eqn:rotcurveterm}; \S\ref{sec:rotcurve}; \S\ref{sec:results_rotcurve}), is significant. Neglecting it is a poor approximation that leads to substantial systematic error (up to a factor $\sim 1.5$) for most wedges (Figs.~\ref{GSE_first_surfdens} and \ref{Sgr600_first_surfdens}). In order to isolate its effect, we measured its functional form directly from the gravitational potential of the simulations. However, this is not directly accessible observationally.

For the real Milky Way, we can estimate $\mathcal{R}$ from the radial Jeans equation \citep{Jeans1915}, which in its most general form, in cylindrical coordinates, is given by \citep{Binney_Tremaine_Book}:

\begin{eqnarray}
\frac{\partial}{\partial R}\left(\nu\mathrm{ \ }\overline{v_R^2} \right) +
\frac{1}{R}\frac{\partial}{\partial \phi}\left(\nu\mathrm{ \ }\overline{v_R v_\phi} \right) +
\frac{\partial}{\partial z}\left(\nu\mathrm{ \ }\overline{v_R v_z} \right) + \nonumber  \\ 
+ \nu\left(\frac{\overline{v_R^2}-\overline{v_\phi ^2}}{R} + \frac{\partial \Phi}{\partial R}\right) 
+ \frac{\partial}{\partial t}\left(\nu \mathrm{ \ }\overline{v_z} \right) 
= 0,
\label{radial_Jeans_eq}
\end{eqnarray}
and $\mathcal{R}$ then follows from $\mathcal{R} = \frac{1}{R}\frac{\partial}{\partial R}\left(R\frac{\partial \Phi}{\partial R}\right) = \frac{1}{R}\frac{\partial v_c^2}{\partial R}$.

The radial Jeans Eq.~(\ref{radial_Jeans_eq}) can be solved locally, similarly to our treatment of the vertical Jeans equation, but it is beyond the scope of this work to explore this in detail. For the real Milky Way, however, we already have some idea of the order of magnitude of $\mathcal{R}$ near the Sun. Using main sequence stars from Gaia DR2, \citet{Li19} find $\frac{\partial v_c^2}{\partial R}(R_\odot) = -1.7 \pm 0.1$\,km\,s$^{-1}$\,kpc$^{-1}$ (and see also \citealt{Ablimit20} and \citealt{eilers19} for similar measurements using different data and methodologies). We can estimate the impact of this on $\rhodm$ from Eq.~(\ref{eqn:rotcurveterm}):

\begin{equation}
    \Delta \rho = \frac{v_c}{2\pi G R}\frac{\partial v_c}{\partial R}
\end{equation}
which, with $R = R_\odot = 8.178 \pm 0.013_{\rm stat.} \pm 0.022_{\rm sys}$\,kpc \citep{gravity19} and $v_c = 229.0 \pm 0.2$\,km\,s$^{-1}$ gives $\Delta \rho = 1.8 \pm 0.1 \times 10^{-3}$\,M$_\odot$\,pc$^{-3}$. This is about a fifth to a tenth of the expected local DM density (Fig.~\ref{fig:rhodmsummary}).

The above estimate of $\Delta \rho$ is only in the Galactic plane and, unfortunately, we need to know its value up to $\sim 2$\,kpc to properly account for it in our analysis (c.f. Fig.~\ref{rot_curve_term_Sgr600}). Selecting wedges in simulations \MWA\ and \MWB\ that have a low $\Delta \rho(z=0)$ (wedges 4, 6 and 7 and \MWA\ and 0, 5 and 7 in \MWB) does not reliably select the best-performing wedges in the absence of the rotation curve term correction (Figs.~\ref{GSE_first_surfdens} and \ref{Sgr600_first_surfdens}). Only wedge 4 in \MWA\ and wedge 5 in \MWB\ are recovered well in the absence of $\mathcal{R}$ because, for these wedges, $\Delta \rho$ remains low at all heights.

Measuring $\Delta \rho(z)$ high above the disc in the Milky Way could be challenging with disc stars due to their rapidly falling density with height. However, halo stars are  excellent for this task. Recently, \citet{Wegg19} have estimated the galactic force field over a substantial volume ($5 < R < 20$\,kpc; $4 < z < 17$\,kpc) around the Sun using RR Lyrae stars in Gaia DR2. This yields a local dark matter density averaged over a larger volume than studies using disc stars. Nonetheless, the value they obtain is in good agreement with more local measurements using the vertical Jeans equation (Fig.~\ref{fig:rhodmsummary} and see  \citealt{Sivertsson:2017rkp} and \citealt{Salomon20}). This may indicate that for our local patch of the Milky Way, the rotation curve term is small at all heights, similarly to wedge 5 of the \MWB\ simulation. It is interesting to note that this wedge is located in an intermediate density region of the disc, next to a more dense spiral arm (Figs.~\ref{stellar_locations} and \ref{Sigma_bary_Sgr600_inc_rotterm_tilt}), similarly to the Sun in the Milky Way \citep[e.g.][]{Eilers20}. Thus, it is possible that we are in a lucky patch of the Galaxy for which $\rhodm$ is not too biased by the rotation curve term.

\subsection{Implications for estimating $\rhodm$ in the Milky Way}

We have found that an accurate estimate of $\rhodm$, in the face of a non-axisymmetric wobbling disc -- like that in the real Milky Way \citep[e.g.][]{Widrow12,Williams13,bovy17,Li19} -- requires careful modelling of the rotation curve and tilt terms (see Eq.~\ref{Jeans_eq}). The axial tilt, 2:nd Poisson and time dependent terms are all much smaller, however, and can be safely neglected, for now.

As noted in prior work, even if the tilt and rotation curve terms can be measured perfectly, an additional problem remains: the degeneracy between $\rhodm$ and the baryonic surface density $\Sigma_b$ \citep[e.g.][]{Sivertsson:2017rkp}. With an error on $\Sigma_b$ of 20\%, comparable to current estimates \citep[e.g.][]{McKee:2015hwa}, we obtained errors on $\rhodm$ of order 30\% for \MWA\ and up to a factor of two for \MWB $\,$ (Fig.~\ref{rhoDM_Sgr600_inc_rotterm_tilt}). We can address this problem in one of three ways. Firstly, we can lower the uncertainty on $\Sigma_b$, which could be reduced by improving our understanding of the local distribution of baryons. Improving the accuracy to $1$\% reduces the formal uncertainty on $\rhodm$ for \MWB\ to $\sim 15$\% (Fig.~\ref{rhoDM_Sgr600_inc_rotterm_tilt}, bottom panel), with model bias error becoming the dominant source of uncertainty. Secondly, we can improve the sampling. \MWA\ has $\sim 4$ times more stars than \MWB, yielding a much smaller uncertainty on $\rhodm$. Finally, we can estimate $\rhodm$ over larger volumes around the Sun for which the dark matter contributes more to the total gravitational potential \citep[e.g.][]{Wegg19, Salomon20}.

Our results suggest, perhaps surprisingly, that time-dependent terms are currently not a leading source of systematic error on $\rhodm$,  for the simulations, number of sample stars and $20$\% uncertainty on $\Sigma_b$ that we have considered as default here.  However, in the future, as more data become available, 
the terms we have found to be unimportant in our current studies will play a larger role.  
With a larger number of sample stars --- leading to a lower statistical uncertainty on $\rhodm$ below $\sim 10$\% --- and/or lower uncertainty on $\Sigma_b$, we find that the time dependent, `axial tilt' and `2:nd Poisson' terms, will all become an important source of systematic uncertainty and will have to also be included in the analysis (see \S\ref{sec:jeans}). 
In the bottom panel of Fig.~\ref{rhoDM_Sgr600_inc_rotterm_tilt}, one can see that if future surveys reduce the errors on $\Sigma_b$ to the $1\%$ level,  we are unable to correctly reproduce $\rhodm$
for wedges 1, 3, 4 and 5 in the \MWB\ simulation
 without adding in the effects of these  terms. The disequilibrium terms could lead to a local systematic error in $\rhodm$ of up to $\sim 30$\%; thus it would be important to add the usually neglected terms to the analysis in order to reduce the uncertainty of $\rhodm$.
Interestingly, these wedges have the largest wedge-to-wedge azimuthal change in the disc surface density (Fig.~\ref{Sigma_bary_Sgr600_inc_rotterm_tilt}), as compared to the mean disc surface density ($\sim 60$M$_\odot$\,pc$^{-2}$). Furthermore, the axial tilt and 2:nd Poisson terms are too small to explain the magnitude of the effect in the bottom panel of Fig.~\ref{rhoDM_Sgr600_inc_rotterm_tilt} (see Fig. 5, which shows corrections from these two terms contribute less than 10\% to $\rhodm$). Hence data with this accuracy in the future will be able to uncover the impact of the time-dependent terms.

\subsection{More realistic mocks}\label{sec:realmocks}
As highlighted from the outset (\S\ref{sec:intro}), our mock data for the Milky Way was set up to be the most optimistic scenario. In reality, we will also have to worry about observational uncertainties, survey selection effects and the potential impact of the Large Magellanic Cloud (LMC; e.g. \citealt{weinberg98,Read:2014qva,gomez15,Laporte18a,Laporte18b,GaravitoCamargo19,deSalas:2020hbh,donaldson21}). We will consider these additional complications in future work, but note here that any impact of the LMC on the Solar Neighbourhood is likely to be sub-dominant to the Sagittarius dwarf that passes much closer \citep[e.g.][]{Laporte19}.

\section{Conclusions}\label{sec:conclusions}

We have assessed the impact of a non-axisymmetric `wobbling' disc on determinations of the local dark matter density 
$\rhodm$. We applied a free-form, steady-state, Jeans method to two different $N$-body simulations of Milky Way-like galaxies. In one, the galaxy experienced an ancient major merger, similar to the hypothesized Gaia-Sausage-Enceladus; in the other, the galaxy was perturbed more recently by the repeated passage and slow merger of a Sagittarius-like dwarf galaxy. We found that common approximations employed in the literature -- axisymmetry and a locally flat rotation curve -- can lead to significant systematic errors of up to a factor $\sim 1.5$ in the recovered surface mass density 
$\sim 2$\,kpc above the disc plane (see Wedge 2 in Figure \ref{GSE_first_surfdens}), implying a fractional error on $\rhodm$ of order unity. However, 
additionally including the tilt term and the rotation curve term
 in our models yielded an unbiased estimate of $\rhodm$, consistent with the true value within our 95\% confidence intervals (Fig.~\ref{rhoDM_Sgr600_inc_rotterm_tilt}) for realistic 20\% uncertainties on the baryonic surface density of the disc. 

For $\sim 30,000$ tracer stars, we obtained an error on $\rhodm$ of a factor of $\sim$two; for $\sim 130,000$ tracers this reduced to $\sim 30$\%. We discussed how to improve on this by increasing the sampling, lowering the uncertainty on the bayronic surface mass density, $\Sigma_b$ and/or estimating $\rhodm$ over larger volumes around the Sun. Reducing the uncertainties on $\Sigma_b$ to 1\% for the simulation with the recent Sagittarius-like merger (with 30,000 tracer stars) lowered the uncertainties on $\rhodm$ to $\sim 15$\%. However, this also led to statistically significant bias on $\rhodm$ of up to 30\% for some locations in the disc. We traced the origin of this bias to the unmodelled time-dependent terms in the Jeans equations.

We conclude that, for an unbiased estimate of $\rhodm$, it is crucial to correctly model the number density distribution of the tracer stars and to include both the `rotation curve' and `tilt' terms (Eq.~\ref{Jeans_eq}). By contrast, the `axial tilt', `2:nd Poisson' and `time dependent' terms are all sub-dominant, even for a non-axisymmetric `wobbling' disc. This leads to the perhaps surprising conclusion that steady-state models are sufficient for estimating $\rhodm$ up to an accuracy of $\sim 30$\%, even in discs that have experienced a recent and significant perturbation. At higher accuracy than this, time dependent terms become important and will need to be included in the models.

\section*{Data Availability}
The data analysed in this article can be made available upon reasonable request to the corresponding author.

\section*{Acknowledgements}
We acknowledge support by the Oskar Klein Centre for Cosmoparticle Physics and Vetenskapsrådet (Swedish Research Council): SS, PFdS, KM and KF through No. 638-2013-8993; AW through No. 621-2014-5772. AW also acknowledges support from the Carlsberg Foundation via a Semper Ardens grant (CF15-0384). KF gratefully acknowledges support from the Jeff and Gail Kodosky Endowed Chair in Physics at the University of Texas, Austin; the U.S. Department of Energy, Office of Science, Office of High Energy Physics program under Award Number DE-SC-0022021 at the University of Texas, Austin; the DoE grant DE- SC007859 at the University of Michigan; and the Leinweber Center for Theoretical Physics at the University of Michigan.This work was supported in part by World Premier International Research Center Initiative (WPI Initiative), MEXT, Japan. CL acknowledges funding from the European Research Council (ERC) under the European Union’s Horizon 2020 research and innovation programme (grant agreement No. 852839).

\bibliographystyle{mnras}
\bibliography{references}
\end{document}